\documentclass[aps,prc,twocolumn,showpacs,superscriptaddress,nofootinbib,floatfix]{revtex4-1}  
\usepackage{graphicx}  
\usepackage{dcolumn}   
\usepackage{bm}        
\usepackage{amssymb}   
\usepackage{amssymb,amsmath,amsfonts} 
\usepackage[breaklinks=true,debug=true]{hyperref}
\usepackage{braket}    
\usepackage{multirow}
\usepackage{adjustbox}
\usepackage{gnuplottex}
\usepackage[T1]{fontenc}
\usepackage[utf8]{inputenc}
\usepackage{color}
\usepackage{dsfont}
\usepackage{tikz}
\usepackage{filecontents}
\usepackage{pgfplots,pgfplotstable}
\usepackage{tikzscale}

\usepgfplotslibrary{groupplots}
\pgfplotsset{compat=newest,
legend image code/.code={
\draw[mark repeat=2,mark phase=2]
plot coordinates {
(0cm,0cm)
(0.15cm,0cm)        
(0.3cm,0cm)         
};%
}
}

\usepgfplotslibrary{units}

\begin{document}


\title{Radial overlap correction to superallowed $0^+ \to 0^+$ nuclear $\beta$ decays using the shell model with Hartree-Fock radial wave functions}

\author{L.~Xayavong}
\affiliation{Physics Department, Faculty of Natural Sciences, National University of Laos, 7322 Dongdok, Vientiane Capital, Lao PDR}
\affiliation{CENBG (CNRS/IN2P3 -- Université de Bordeaux), 33175 Gradignan cedex, France}
\author{N.~A.~Smirnova} 
\affiliation{CENBG (CNRS/IN2P3 -- Université de Bordeaux), 33175 Gradignan cedex, France}
\vskip 0.25cm  
\date{\today}

\begin{abstract} 
The radial overlap correction, $\delta_{RO}$ is re-examined for 30 superallowed $0^+\to 0^+$ nuclear $\beta$ decays using the shell model with Hartree-Fock (HF) radial wave functions. Our mean-field calculation is based on the effective Skyrme interaction including Coulomb, Charge-Symmetry-Breaking (CSB) and Charge-Independence-Breaking (CIB) terms. In addition, the electromagnetic corrections such as those due to the gradient of charge density, vacuum polarization, Coulomb spin-orbit and finite size of nucleons are also considered. 
In order to avoid the spurious isospin mixing, the local equivalent potential is constructed from the solution of a HF calculation for the $Z=N$ nucleus with charge-dependent forces neglected. Then the obtained mean field is solved non-iteratively for the parent and daughter nuclei. 
It turns out that the CIB term has no significant impact on $\delta_{RO}$ throughout the mass range between 10 and 74. On the other hand, the CSB term makes $\delta_{RO}$ increasing systematically between 10 and 30~\%. Similarly, the gradient density leads to a further increase of $\delta_{RO}$ between 2 and 14~\%, while other estimated electromagnetic corrections are negligible. The effect of the suppression of isospin spuriosity is somewhat complicated. In general, it produces a significantly larger value of $\delta_{RO}$, however there are few cases for which $\delta_{RO}$ is mostly unaffected or even reduced, especially the even-even emitters in the light mass region. 
All these improvements partly explain the long-standing discrepancy between the correction values obtained 
with Woods-Saxon (WS) and HF radial wave functions. Nevertheless, the remaining discrepancy is still significant, except for the emitters with $A\le 26$. In addition, the local odd-even staggering presented on the results obtained with the phenomenological potential~\cite{HaTo2015,HaTo2015x,XaNa2018} are not well reproduced by our calculations. This subsequently leads to a large difference in the predicted mirror $ft$ ratios. These problems might relate to the errors of the calculated charge radii as well as the deformation and other correlation effects on the data of separation energies used for constraining the asymptotic radial wave functions.
\end{abstract} 

\maketitle

\section{Introduction}

It has been pointed out that the superallowed nuclear $\beta$ decays between $(J^\Pi=0^+,T=1)$\footnote{Throughout this article we use the capital $\Pi$ for parity and the small $\pi$ for intermediate state label which will be specified later.} states provide an excellent tool to probe the properties of the electroweak interaction and have been the subject of intense study for several decades~\cite{ToHa1977,f,ToHaRL1994,ToHa2002,ToHa2008,HaTo2009,ToHaRPP2010,ToHa2010,HaTo2015,HaTo2015x}. As an interesting feature, the wave function of the entire nucleus is left unchanged during the transition, except for one proton being converted into a neutron or vice-versa. Thus, if the initial and final states are perfect analogue states, the nuclear matrix element of this transition is model-independent and the value for the vector coupling constant, $G_V$ could be extracted from measured $ft$ values. The most accurately determined $ft$ values~\cite{HaTo2015,Ben2019}, however, are not constant within experimental uncertainty, suggesting the breaking of analogue symmetry between the initial and final nuclei as well as the radiative effects such as bremsstrahlung. Specifically, $G_V$ is obtained from each $ft$ value via the relation~\cite{HaTo2015}, 
\begin{equation}\label{Ft}
\mathcal{F}t = ft(1+\delta_R')(1+\delta_{NS}-\delta_C) = \frac{K}{2G_V^2(1+\Delta_R^V)}, 
\end{equation}
where $K$ is a combination of fundamental constants, $K/(\hbar c)^6 = 2\pi^3\ln(2)\hbar/(m_ec)^5 = 8.120270(12)\times 10^{-7}~\mbox{GeV}^{-4} \mbox{sec.}$, $ft$ is the product of the statistical rate function ($f$)~\cite{f} and the partial half-life ($t$). The radiative corrections are separated into three parts~\cite{HaTo2015} : 
$\Delta_R^V= 2.467(22)\%$ is nucleus independent~\cite{Chien2018}, 
$\delta_R'$ depends on the atomic number of daughter nucleus and 
$\delta_{NS}$ is nuclear-structure dependent. The correction due to the isospin-symmetry breaking, $\delta_C$, is defined as the deviation of 
the realistic Fermi matrix element squared from its isospin-symmetry limit value squared, $|M_F^0|^2$, such as, 
\begin{equation}\label{mf}
|M_F|^2 = |M_F^0|^2(1-\delta_C),
\end{equation}
where $|M_F^0|=\sqrt{T(T+1)-T_{zi}T_{zf}}$, $T$ is the isospin quantum number for an isobaric multiplet, $T_{zi}$ and $T_{zf}$
are isospin projection quantum numbers of the initial and final states, respectively. 

The constancy of the corrected $\mathcal{F}t$ value serve as a direct test of the Conserved Vector Current (CVC) hypothesis. 
More importantly, if CVC holds, the comparison between the vector coupling constant for the purely leptonic $\mu$ decay, $G_\mu$~\cite{MuLan} and the semi-leptonic nuclear Fermi $\beta$ decay, $G_V$, will yield the $|V_{ud}|$ matrix element of the Cabbibo-Kobayashi-Maskawa (CKM) quark-mixing matrix as 
$$
|V_{ud}|=G_V/G_{\mu }\,.
$$ 

Subsequently, a precise test of the unitarity condition of the CKM matrix under the assumption of the three-generation Standard Model is possible~\cite{HaTo2015}. Although the role played by nuclear structure is relatively small, the precision reached by experiment  nowadays is such that theoretical uncertainties introduced by correction terms required in the analysis of the $ft$-value data
predominate over the experimental uncertainties. Two of these correction terms ($\delta_C$ and $\delta_{NS}$) depend on nuclear structure, and together they have been identified as the largest contributor to the overall uncertainty in the corrected $\mathcal{F}t$ values~\cite{HaTo2015}. Therefore, an improvement of these corrections, especially $\delta_C$ which dramatically depends on nuclear structure models~\cite{XaNa2018,ToHa2010}, is highly demanded. 

In spite of great progress in microscopic many-body calculations, precise description of the isospin-symmetry breaking is still a
challenge. Unlike the other nuclear properties, the isospin mixing is induced mainly by the long-range Coulomb force, implying that the diagonalization of an isospin non-conserving (INC) Hamiltonian has to be performed in a model space which includes several oscillator shells~\cite{nocore,NaXa2018}. However, instead of doing a first-principle shell-model calculation, one can approximately separate the isospin-breaking correction into two components~\cite{OrBr1985,XaNa2018,ToHa2002}, 
\begin{equation}
\delta_{C}\approx\delta_{RO}+\delta_{IM}, 
\end{equation}
where $\delta_{IM}$ accounts for the isospin admixture within the modest shell-model space, while $\delta_{RO}$ accounts for the mismatch between proton and neutron radial wave functions, simulating the admixture between states that lie outside the shell-model configuration space. 
The isospin-mixing component, $\delta_{IM}$, is generally very sensitive to the effective INC interaction,
because of the strong dependence on the energie difference between admixed states (analogue and non-analogue $0^+$ states). 
Fortunately, the contribution of $\delta_{IM}$ to the total, $\delta_C$ correction is normally less than 10~\%~\cite{NaXa2018}. 
As a traditional technique to improve this undesired property, the calculated $\delta_{IM}$ values are scaled 
with the measured energy separation between the first and second $0^+$ states in daughter nucleus. 
Althought this technique is based on the limit of two-level mixing, it works adequately for many cases~\cite{ToHa2008}. 
On the other hand, the study of the larger component, $\delta_{RO}$, cannot be satisfactorily concluded because the values calculated by Towner and Hardy (TH) using WS radial wave functions~\cite{ToHa2008,HaTo2015,HaTo2015x} yield considerably different $\mathcal{F}t$ values than do those of the evaluation of Ormand and Brown (OB) using Skyrme HF radial wave functions~\cite{OrBr1985,OrBr1989x,OrBr1995,OrBr1989}. 
This discrepancy has been thought to be partially due to the use of the Slater approximation for treating the Coulomb exchange term. 
In particular, it has been shown that the resulting Coulomb potential is too extensive or overestimated at large distance~\cite{HaTo2009}. 
Unfortunately, in addition to the numerical inconvenience, the exact treatment of the Coulomb exchange term generates a nonlocal component to the Skyrme HF mean field~\cite{Skalski} and then is technically unsuitable to be implemented within the OB's protocol which incorporates the multiple intermediate $(A-1)$-particle states~\cite{OrBr1985}. 
As an asymptotic correction, Towner and Hardy proposed to replace the HF potential of the parent and daughter nuclei identically with the HF potential of the intermediate $(A-1)$ nucleus, the calculation based on this modified protocol yields the $\delta_{RO}$ values which are very close to those obtained with WS radial wave functions, except that their local variations (odd-even staggering) trend to occur in the opposite direction~\cite{HaTo2009}. We remark that, although this method ensures the correctness of the Coulomb potential at large distance, it may disturb or even destroy the other important properties of the Skyrme HF mean field at small to medium distances. Furthermore, the dependence of $\delta_{RO}$ on the choice of the Skyrme force parametrizations has been studied in Refs.~\cite{Lat,Xthesis}, this effect was found to be quite small comparing with the gaps between the OBs and THs values. 

It is the aim of this work to extend the OB's study of $\delta_{RO}$ in several directions. In addition to the update of the shell-model effective interactions and configuration spaces as discussed in section~\ref{sm}, we propose an empirical scheme to avoid the spurious isospin mixing in a self-consistent HF calculation. Our main idea is to construct a realistic spuriosity-free potential which, within the effective Skyrme interaction, is a known functional of densities~\cite{karim}, from an isospin-invariant HF solution of the nearby $N=Z$ isobar. More details on this step are described in subsection~\ref{fit}. 
We also study two different approaches to the center-of-mass correction which give rise to a difference in mass and, in turn, directly affect the asymptotic radial wave functions. One of them is the one-body correction term commonly used for the mean-field calculations within the effective Skyrme interaction~\cite{Vautherin}, and the other one is based on the concept of reduced mass commonly used with the phenomenological WS potential~\cite{XaNa2018,SWV,BohrMott}. This study is given in subsection~\ref{ccom}. 
In order the check the validity of the Slater approximation, the calculation of $\delta_{RO}$ using the exact Coulomb exchange term has been carried out in the closure approximation~\cite{XaNa2018,Xthesis} and without re-adjustment of the HF potential, the results are discussed with special attention paid to the deviation of the calculated displacement energy between the initial and final states from the experimental data. More details on this calculation are given in section~\ref{EM}. Furthermore, the corrections to the Coulomb potential due to the effects of charge density gradient, vacuum polarization, Coulomb spin-orbit and finite size of nucleons which were found to have a non-negligible contribution to the isospin-symmetry-breaking properties~\cite{Roca-Maza,Naito1,Naito2,Naito} are also explored. 
At the interaction level, we consider two extended isospin non-conserving forces complemented to the conventional effective Skyrme interaction, namely the charge-symmetry-breaking (CSB) and charge-independence-breaking (CIB) forces as detailed in Ref.~\cite{Sagawa}. Their contribution to the HF potential is derived in section~\ref{isb}. 
Section~\ref{result} is devoted to a presentation and a discussion of our final results as well as their impact on the Standard Model. Some general conclusions are drawn and perspectives are suggested in section~\ref{con}.

\section{$\delta_{RO}$ in the parentage expansion formalism}\label{forma} 

The initial and final state vectors obtained with the shell model are given in terms of the multi-particle angular structure in
the model space and do not explicitly involve the radial wave functions. 
In order to apply the configuration-mixing results to the calculation of one-body transition densities, such as those for the isospin-raising/lowering operators, $\hat{T}_{\pm}$ in $\beta$ decays, an appropriate set of radial wave functions must
be separately introduced. For the superallowed $0^+\to 0^+$ nuclear $\beta$ decays, if one uses a realistic single-particle potential, such as WS or self-consistent HF instead of the usual harmonic oscillator, the radial wave functions of protons in the decaying nucleus will be pushed out relative to those of neutrons in the daughter nucleus. Such a radial mismatch leads to an extra reduction in modulus of the Fermi matrix element from its model-independent value~\cite{XaNa2018,ToHa2008} which, by definition, is accounted for in the $\delta_{RO}$ component of the isospin-breaking correction. This effect is basically due to the Coulomb and charge-dependent forces of nuclear origin, as well as the isovector potential\footnote{The mean-field potential derived with a Skyrme-HF calculation can be decomposed into an isoscalar, an isovector (or (a-)symmetry term) and Coulomb potentials~\cite{DG}. The isovector term makes the nuclear part of the proton field to be more attractive and inversely for the neutron field.} induced by the difference in the proton and neutron densities of a single nucleus~\cite{OrBr1985}. 

By expanding the $A$-particle wave functions $\ket{i}$ and $\ket{f}$ into products of $(A-1)$-particle wave functions $\ket{\pi}$ and single-particle functions $\ket{\alpha}$, the correction $\delta_{RO}$ can be expressed in the following $JT$-coupled form,
\begin{equation}\label{ro}
\displaystyle\delta_{RO} = \frac{2\hat{J}_f}{M_F^0}\sum_{k_\alpha\pi} C^p_{T_\pi}C^n_{T_\pi} S_{k_\alpha,\pi}^f(1-\Omega_{k_\alpha}^\pi),
\end{equation}
where $\hat{J}_f=\sqrt{2J_f+1}$ and $k_\alpha$ is shorthand for the set of spherical quantum numbers $(nlj)$ of state $\alpha$. The sum in Eq.~\eqref{ro} is over all intermediate states $\ket{\pi}$ and all single-particle orbitals $\ket{\alpha}$ active in the shell-model computation. The isospin Clebsch-Gordan coefficient, $C^{t_z}_{T_\pi}=\braket{T_\pi T_{\pi z}tt_z|T_fT_{fz}}$, is obtained analytically using the formula,  
\begin{equation}\label{cg}
\begin{array}{l}
\displaystyle C^{t_z}_{T_\pi} =\pm\sqrt{\frac{1}{2}\left(1\pm2t_z\frac{T_{\pi z}+t_z}{T_\pi+t}\right)}, 
\end{array}
\end{equation} 
where $t=1/2$ and $t_z=1/2(-1/2)$ for neutron(proton). The $\pm$ signs correspond to $T_\pi=T_f\mp 1/2$, and the one in front of the square root must be omitted if $t_z=-1/2$. 

The spectroscopic factor, $S_{k_\alpha,\pi}^f$ is related to the overlap of the $A$ and $(A-1)$-particle wave functions. Its definition is 
\begin{equation}\label{sf}
\displaystyle S_{k_\alpha,\pi}^f = \left| \frac{ \braket{f||| a_{k_\alpha}^\dagger |||\pi} }{\hat{J}_f\hat{T}_f} \right|^2, 
\end{equation}
here $\hat{T}_f$ is defined analogously as $\hat{J}_f$. The triple bar matrix element in Eq.~\eqref{sf} indicates that it is reduced 
in both coordinate space and isospace. 
Since the initial $\ket{i}$ and final $\ket{f}$ states are isobaric analogue, and the isospin mixing is not included in the calculation of $\delta_{RO}$, the spectroscopic factor~\eqref{sf} is invariant by changing neutron to proton or vice-versa, that is $S_{k_\alpha,\pi}^f=S_{k_\alpha,\pi}^i$. 

We remark also that Eq.~\eqref{ro} links the radial overlap correction to the spectroscopic factors which, in principle, could be measured using the single-nucleon transfer reactions~\cite{Pr2011,Ly1967,Fu1985}. Although the spectroscopic factor data for proton-rich nuclei around the $N=Z$ line are very limited and imprecise, they have been efficiently used as a guide for the model-space selection, especially for nuclei in the cross-shell regions~\cite{ToHa2008}. More detail on this will be discussed in section~\ref{sm}. 

The symbol $\Omega_{k_\alpha}^\pi$ denotes the overlap integral of radial wave functions between the decaying proton, $R_{k_\alpha,p}^\pi(r)$ and the resulting neutron, $R_{k_\alpha,n}^\pi(r)$. It is written as, 
\begin{equation}\label{overlap}
\displaystyle \Omega_{k_\alpha}^\pi = \int_0^\infty R_{k_\alpha,p}^\pi(r)R_{k_\alpha,n}^\pi(r)r^2 dr < 1. 
\end{equation} 

Obviously, if $R_{k_\alpha,p}^\pi(r)$ and $R_{k_\alpha,n}^\pi(r)$ are identical, $\Omega_{k_\alpha}^\pi$ will be reduced to the normalization integral. 
Otherwise, $\Omega_{k_\alpha}^\pi$ will be slightly less than unity and leads to a non-vanishing $\delta_{RO}$ component. 
Therefore, a large contribution to $\delta_{RO}$ requires a large spectroscopic factor and a significant departure of the overlap integral~\eqref{overlap} from unity. 

It should be noted that, within realistic single-particle basis, Fermi transitions between orbitals with different numbers of nodes are not strictly forbidden. For example, the overlap integral between neutron in $1s_{1/2}$ and proton in $2s_{1/2}$ may not be vanished perfectly. Nevertheless, such a nodal mixing effect cannot be taken into account in a straightforward manner due to the requirement of a huge model space. A generalized formalism for $\delta_C$ has been discussed by Miller and Schwenk~\cite{MiSch2008,MiSch2009}.

Next, we substitute $T_f = 1$ and separately identify the sums to the isospin-lesser states
with $T_\pi = 1/2$, denoted $\pi^<$, and the isospin-greater states with
$T_\pi = 3/2$, denoted $\pi^>$, so that Eq.~\eqref{ro} can be decomposed as follows  
\begin{equation}\label{ro1}
\begin{array}{l}
\displaystyle\delta_{RO} = \sum_{k_\alpha\pi^<} S_{k_\alpha,\pi^<}^f(1-\Omega_{k_\alpha}^{\pi^<}) - \frac{1}{2}\sum_{k_\alpha\pi^>} S_{k_\alpha,\pi^>}^f(1-\Omega_{k_\alpha}^{\pi^>}). 
\end{array}
\end{equation}

Eq.~\eqref{ro1} demonstrates that there is always a cancellation between the contributions of the isospin-lesser states and
the isospin-greater states. In particular, this effect would be more substantial for deeply bound orbitals for which the French-MacFarlane sum rule~\cite{ToHa2008,FrMac1961} yields 
\begin{equation}\label{fm}
\displaystyle \sum_{\pi^<} S_{k_\alpha,\pi^<}^f\approx\frac{1}{2}\sum_{\pi^>} S_{k_\alpha,\pi^>}^f. 
\end{equation}
See section~\ref{sm} for further details on this point.

The label $\pi$ in Eq.~\eqref{overlap} means that the neutron and proton wave functions are taken as eigenfunctions of a realistic potential whose depth is adjusted to reproduce
the neutron or proton separation energies in the respective nucleus which, in turn, depend on the intermediate $(A-1)$-particle state $\ket{\pi}$. 
For example, if the intermediate state is the
ground state of the $(A-1)$ system, then the proton separation energy would be $S_p$ and the neutron separation energy $S_n$, which are well-known from experiment and can be found in any atomic mass table. If, however, $\pi$ represents an excited state of the $(A-1)$ system, then the proton and neutron separation
energies would be $S_p+E_\pi^x$ and $S_n+E_\pi^x$, respectively, where $E_\pi^x$ is the excitation energy of that intermediate state. 
If we do not allow the proton and neutron radial functions $R_{k_\alpha,p}^\pi(r)$ and $R_{k_\alpha,n}^\pi(r)$ to vary with the states $\pi$ but fix their asymptotic forms for all $\alpha$ to the separation energies of the ground state of the $(A-1)$ system, 
then the sums over $\pi$ can be done analytically and the computed value of $\delta_{RO}$ 
becomes independent of the shell-model effective interaction (see for example Eq.~(18) of Ref.~\cite{ToHa2008}). 

\begin{figure}[htb]
\begin{center}
\includegraphics[]{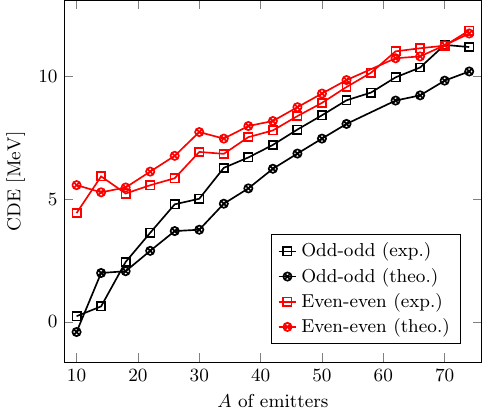}
 \caption{(Color online) Coulomb displacement energy between the initial and final states with $T=1$ and $J^\Pi=0^+$ in function of mass number $A$. The theoretical values (theo.) were obtained from a spherical HF calculation using the zero-range Skyrme interaction~\cite{skm*}. The experimental data (exp.) were taken from the AME2012~\cite{AME2012}.}
\label{figx}
\end{center}
\end{figure}

It is worth noticing here that the fit single-particle energies to nucleon separation energies provides a direct constraint on the overlap integral~\eqref{overlap}, 
since the asymptotic form of radial wave functions depends on the value of the single-particle energy $\epsilon $ roughly as 
$$
\displaystyle R(r) \to \exp{\left( -\frac{\sqrt{2m |\epsilon |}r}{\hbar}\right)}. 
$$
Furthermore, within the potential-depth prescription just mentioned, the magnitude of $\delta_{RO}$ is principally determined by the difference between $S_n$ and $S_p$ which is equal to the Coulomb Displacement Energy ($CDE$) between the initial and the final states, 
\begin{equation}\label{cde}
\begin{array}{ll}
CDE &= \displaystyle E_i(0^+,T=1)-E_f(0^+,T=1) \\[0.15in]
    &= \displaystyle \left[ E_i(0^+,T=1)-E_\pi(J^\Pi_\pi,T_\pi) \right] \\[0.15in]
    &- \displaystyle \left[ E_f(0^+,T=1) - E_\pi(J^\Pi_\pi,T_\pi) \right] \\[0.15in]
    &= \displaystyle -S_p+S_n, 
\end{array}
\end{equation}
where $E_\pi(J^\Pi_\pi,T_\pi)$ is the energy of the intermediate state with spin-parity $J^\Pi_\pi$ and isospin $T_\pi$. It can be separated into two terms, namely $E_\pi(J^\Pi_\pi,T_\pi)=E_\pi(gs.)+E_\pi^x(J^\Pi_\pi,T_\pi)$ where $E_\pi(gs.)$ is the ground-state energy and $E_\pi^x$ the excitation energy as mentioned in the previous paragraph. Note that if the initial or final state is not ground state, its excitation energy must also be introduced to Eq.~\eqref{cde}. 

As illustrated in Fig.~\ref{figx}, the $CDE$ increases almost linearly with the mass number, except for the presence of local fluctuations (odd-even staggering) which can be seen over the mass range of interest. The local effect is such that, for a given mass, the $CDE$ value of the even-even emitter is larger than that of the odd-odd emitter. In particular, the local difference in $CDE$ becomes more substantial for the light nuclei with $A\le 18$. 
Although it is impossible to extract $\delta_{RO}$ directly from these experimental data, 
one can expect them to share some common features with the correction, in particular the local odd-even staggering. 
Nevertheless, our theoretical prediction based on the Skyrme-HF approximation differs significantly from the measured $CDE$ values, especially for the odd-odd emitters with $A\ge26$. Therefore, if this discrepancy is dominated by a post-HF effect, a fine-tuning of the self-consistent mean field would be ambiguous and may potentially deteriorate the agreement of the resulting correction values with the Standard Model' hypotheses. Further discussion is given in section~\ref{result}. 

\section{Shell model calculations}\label{sm} 

\begin{figure}[htb]
\begin{center}
\includegraphics[]{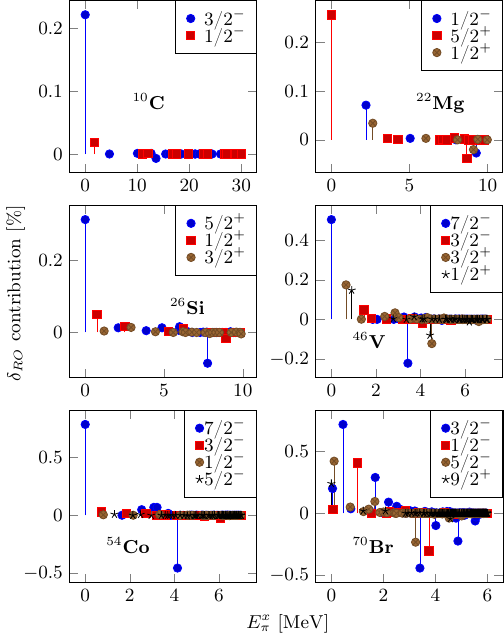}
\caption{(Color online) Distribution of $\delta_{RO}$ for a selected transition from each model space as a function of the intermediate state excitation energy in a corresponding $(A-1)$ nucleus. 
Vertical line with symbols on top show individual contributions to $\delta_{RO}$ from intermediate states. 
By convention, states with isospin $T_\pi=1/2$ provide a positive contribution, which correspond to the first term of 
Eq.~\protect\eqref{ro1}, while $T_\pi=3/2$ states contribute with the opposite sign as given by the second term of Eq.~\eqref{ro1}.}
\label{fig2}
\end{center}
\end{figure}

The calculation of $\delta_{RO}$ using the full parentage formalism (Eq.~\eqref{ro1}) involves the sum over the complete set of intermediate $(A-1)$-particle states for which a robust cut-off must be imposed at a certain energy level. In general, the contribution of high-lying states (typically above 10~MeV) can be neglected since the spectroscopic factor decreases on average with increasing  excitation energy. At the same time, the excitation of the intermediate $(A-1)$ system makes, respectively, the decaying proton and resulting neutron to be more bound to the parent and daughter nuclei. Therefore, the mismatch of the respective radial wave functions is subsequently reduced. The variation of $\delta_{RO}$ as a function of excitation energy for some selected transitions is shown in Fig.~\ref{fig2}. As a reasonable cut-off for the sum, our calculations include a number of lowest intermediate $(A-1)$-particle states for a given spin-parity in the range from a dozen to 350, depending on the size of the configuration space. 

Based on the energy-dependent behavior of the overlap integral and the fact that all isospin-greater states lie above the spectroscopically significant isospin-lesser states, one immediately deduces that $\Omega_{k_\alpha}^{\pi^<}$ deviates more significantly from unity than $\Omega_{k_\alpha}^{\pi^>}$. Therefore, the cancellation between the first and second terms of Eq.~\eqref{ro1} is only partial. 
This effect leads to a significant and systematic deviation of the corresponding calculated $\delta_{RO}$ values 
from those obtained within the description using the traditional closure approximation~\cite{XaNa2018}. 
In particular, it leads to a considerable amplification of the closed-shell contribution which, according to the French-MacFarlane sum rule~\eqref{fm}, should be negligible if there is no difference between $\Omega_{k_\alpha}^{\pi^<}$ and $\Omega_{k_\alpha}^{\pi^>}$. 
This property has been discussed by Towner and Hardy in their earlier works~\cite{ToHa2002,ToHa2008}, and more recently re-examined by Xayavong and Smirnova~\cite{Xthesis,XaNa2018}. 

To account for this induced core polarization, the calculation of $\delta_{RO}$ would generally require a larger configuration space comparing with the calculation of the isospin-mixing $\delta_{IM}$ part. As a well-known example, it has been shown~\cite{ToHa2008} for $^{46}$V$\to^{46}$Ti that the $pf$ shell is not sufficient for the calculation of $\delta_{RO}$. Although this model space generated reasonable energies and spins for the known states in $^{46}$V and $^{46}$Ti, it failed to produce the low-lying intruder states in $^{45}$Ti which are strongly populated in single-nucleon pick-up reactions such as $(p,d)$ and ($^3$He,$\alpha$)~\cite{Borlin}, especially the $3/2^+$ state at an excitation energy of only $330$~keV. Such $sd$-shell states can contribute to 
the structure of the initial and final states of the superallowed transition and consequently must affect the radial mismatch between them. 
This indicates that a complete calculation of the radial overlap correction for the decay of $^{46}$V should include contributions from deeper lying $sd$-shell orbitals. 
Our predicted contributions of $1d_{3/2}$ and $2s_{1/2}$ to $\delta_{RO}$ for the mentioned transition are displayed in the middle-right panel of Fig.~\ref{fig2}. In fact, the polarization of the $sd$-core seems to be dominated across the lower $pf$ shell up to $A=50$~\cite{SFN1979,Ly1967,Pr2011,Fu1985}. 
However, diagonalization of a Hamiltonian matrix in the full $sd$-$pf$ shell-model space could not be done easily, 
because of the computational limitations. 
Another difficulty arises from a possible contribution of the center-of-mass motion in the extended model space. 

Besides, a specific computational challenge is the shell-model description of the heavy-deformed nuclei in the upper
 $pf$ shell with $A\ge 62$. Since the conventional shell model uses a spherical basis, a large configuration space is normally needed 
to produce deformation degrees of freedom. 

In this work, large-scale diagonalizations have been performed using the shell-model code NuShellX@MSU~\cite{NuShellX}. 
We considered only one model space and one effective interaction for a given transition; the uncertainty arising from the use of different interactions has already been explored by several authors~\cite{Xthesis,ToHa2002,ToHa2008}. We used the Cohen-Kurath interaction~\cite{Cohen1967} for nuclei with mass between $10$ and $14$, diagonalizing it in $1p$ shell. 
To ensure that the polarization of the $1p$ core is properly accounted, the interaction of McGrory-Wildenthal~\cite{MW1973} with the model space comprising of $1p_{1/2}1d_{5/2}2s_{1/2}$ was used for $18\le A\le 22$. The well-known universal $2s1d$-shell interaction~\cite{USD} was considered for $26\le A\le 34$. 
In fact, it was experimentally found that there exist negative parity states in the spectra of $A=25$ and $26$ nuclei, carrying significant spectroscopic strength from the $1p$ shell~\cite{Joyce,Peterson}.
However, most of these states lie above 10~MeV, therefore their contribution to $\delta_{RO}$ should be small and is thus neglected for the present study. 
The nuclei in the vicinity of $^{40}$Ca, including those with $38\le A\le 46$, were calculated in the model space spanned by 
$2s_{1/2}1d_{3/2}1f_{7/2}2p_{3/2}$ orbitals using the modified version of the ZBM2 interaction~\cite{Bis2014}. 
Nuclei with $50\le A\le 54$ are too heavy for $2s1d$-$2p1f$ model space without truncation, 
we therefore adopted for them the traditional $2p1f$ shell using the GXPF1A~\cite{gx1a} interaction. 
While for $62\le A\le 74$, we employed the JUN45 interaction~\cite{jun45} in the model space formed by 
$2p1f_{5/2}1g_{9/2}$ orbitals. No further modification was made to the chosen effective interactions. 

Note that Towner and Hardy~\cite{ToHa2008,HaTo2015} also included $1d_{5/2}$ to the model space for $A=38,42$, and included $2s_{1/2}1d_{3/2}$ for $A=50,54$. We agree that contributions of these orbitals may be non-negligible for these nuclei. However, such a huge configuration space cannot be treated without truncation. We therefore keep it for the future development of computational resources. Our main focus here is rather the comparative study of the various terms at the mean-field level aimed at clarifying the discrepancy caused by the use of different potential types. 

\section{Evaluation of the overlap integrals using Hartree-Fock radial wave functions}\label{hf}
\subsection{Skyrme-Hartree-Fock Potential}\label{hfpo}

The HF method has been extensively explored as a fundamental means of generating the average nuclear field. The HF calculations with an effective force due to Skyrme~\cite{Skyrme1959} have been particularly successful in describing a number of bound-state properties of spherical closed-shell nuclei, in particular the charge radius, total binding energy and level density~\cite{Vautherin,karim,Bender}. 
The derivation of the HF equations corresponding to the conventional Skyrme interaction has been discussed by Vautherin and Brink~\cite{Vautherin}. Here we only recall the basic results which are relevant for our purposes. 

Restricting to the spherical and time reversal symmetries, the HF calculations using the standard form of Skyrme interaction (see appendix~\ref{appA}) lead to a set of nonlocal differential equations with eigenfunctions $R_{\alpha,q}(r)= u_{\alpha,q}(r)/r$ and eigenvalues $\epsilon_{\alpha,q}$, 
where $q=p$ for protons and $q=n$ for neutrons. 
The nonlocal eigenfunctions and eigenvalues can be obtained from the following equivalent set of differential equations, which involve a local energy-dependent potential~\cite{DG}, 
\begin{equation}\label{equiv}
\displaystyle-\frac{\hbar^2}{2m}\frac{d^2}{dr^2} u_{\alpha,q}^L(r) + \tilde{U}_{\alpha,q}^L(r)u_{\alpha,q}^L(r) = \epsilon_{\alpha,q} u_{\alpha,q}^L(r), 
\end{equation}
where $u_{\alpha,q}(r)$ and $u_{\alpha,q}^L(r)$ are connected by a damping factor: 
\begin{equation}\label{u}
\displaystyle u_{\alpha,q}(r)=N_q\left[\frac{m_q^*(r)}{m}\right]^{1/2} u_{\alpha,q}^L(r), 
\end{equation}
with $N_q$ being the normalization constant. In principle, one may solve the HF equation directly in its standard form~\cite{karim}. 
The merit of the introduction of $\tilde{U}_{\alpha,q}^L(r)$ is to facilitate the comparison with results of the phenomenological WS calculations~\cite{ToHa2008,HaTo2015}. The mass $m$ is taken as the average between the proton and neutron mass. We have checked that the mass difference between neutron and proton is too small to produce any visible effect on the $\delta_{RO}$ correction. 

The local equivalent potential, $\tilde{U}_{\alpha,q}^L(r)$ including the centrifugal barrier term, can be decomposed as~\cite{DG,Xthesis}
\begin{equation}\label{local}
\begin{array}{ll}
\displaystyle \tilde{U}_{\alpha,q}^L(r) &= \displaystyle\frac{\hbar^2}{2m}\frac{ \braket{\boldsymbol{l}^2} }{r^2} + \frac{m_q^*(r)}{m} \Bigg\{ U_q(r) +\frac{1}{4} \frac{d^2}{dr^2} \frac{\hbar^2}{m_q^*(r)} \\[0.18in]
&- \displaystyle\frac{m_q^*(r)}{2\hbar^2}\left[ \frac{d}{dr} \frac{\hbar^2}{m_q^*(r)} \right]^2 + \frac{1}{2}W_q(r)\braket{\boldsymbol{\sigma}\cdot\boldsymbol{l}} \\[0.18in]  
&+ \displaystyle \delta_{qp}V_{Coul}(r) \Bigg\} + \left[ 1-\frac{m_q^*(r)}{m} \right]\epsilon_{\alpha,q},  
\end{array}
\end{equation}
where $\delta_{qp}=1(0)$ for $q=p(n)$. The expressions for the central $U_q(r)$, spin-orbit $W_q(r)$ and Coulomb $V_{Coul}(r)$ potentials as well as for their respective energy density functionals and for the effective mass $m_q^*(r)$ are given in appendix~\ref{appA} (subsections~\ref{a2} and \ref{a3}.

The nucleon densities can be expressed in term of the local eigenfunctions: 
\begin{equation}
\rho_q(r) = \frac{N_q^2}{4\pi r^2}\sum_\alpha \nu_\alpha^q \left[\frac{m_q^*(r)}{m}\right] \left|u_{\alpha,q}^L(r)\right|^2, 
\end{equation}
where $\nu_\alpha^q$ is the occupation number of the corresponding state $\alpha,q$. In order to restore the breaking of the time-reversal symmetry in an odd-mass nucleus, we use the so-called equal-filling approximation for occupation numbers. Likewise, the spin-current densities are given as, 
\begin{equation}
\displaystyle 
J_q(r) = \frac{N_q^2}{4\pi r^3}\sum_\alpha \nu_\alpha^q \braket{\boldsymbol{\sigma}\cdot\boldsymbol{l}} 
         \left[\frac{m_q^*(r)}{m}\right]\left|u_{\alpha,q}^L(r)\right|^2, 
\end{equation}
and the kinetic densities as, 
\begin{equation}
\begin{array}{ll}
\tau_q(r) &= \displaystyle\frac{N_q^2}{4\pi r^2}\sum_\alpha \nu_\alpha^q \left[\frac{m_q^*(r)}{m}\right] \\[0.18in] 
          &\times \displaystyle\left\{\left[ \frac{d}{dr}u_{\alpha,q}^L(r)-\frac{u_{\alpha,q}^L(r)}{r} \right]^2+\frac{ \braket{\boldsymbol{l}^2} }{r^2}\left|u_{\alpha,q}^L(r)\right|^2\right\}, 
\end{array}
\end{equation}
whereas the total densities (without index) are given by $\rho(r) = \rho_p(r) + \rho_n(r)$, $J(r)=J_p(r)+J_n(r)$ and $\tau(r)=\tau_p(r)+\tau_n(r)$. 

The deviation of the effective mass from the bare nucleon mass is a measure of non-locality of the Skyrme-HF mean field and seems to be an obvious source of the difference between the self-consistent HF and the phenomenological WS potential~\cite{DG}. 
In particular, the local equivalent potential, $\tilde{U}_{\alpha,q}^L(r)$ comprises an energy-dependent and two derivative terms whose magnitude is determined by the deviation of ${m_q^*(r)}/m$ from unity and the spatial variation of $m_q^*(r)$, respectively. 
This nuclear medium effect manifests itself mainly in the nucleus interior and vanishes asymptotically, such that
\begin{equation}\label{asymp}
\begin{array}{lll}
u_{\alpha,q}(r)=u_{\alpha,q}^L(r), & & \text{as} \hspace{0.1in} r\to \infty. 
\end{array}
\end{equation}  

On the other hand, if the effective mass is constant, the damping factor in Eq.~\eqref{u} will be removed by the normalization of radial wave functions. In this case, Eq.~\eqref{asymp} also holds for any value of $r$, 
regardless the difference between $m^*$ and $m$. 

As a functional of the different densities, the nuclear part of the Skyrme HF field generally depends on the isospin orientation. 
It was demonstrated that this property is analogous with the (a-)symmetry term of the phenomenological WS potential~\cite{DG}, 
except for the spurious isospin mixing contribution arising from the variational principle and not from the forces. 
The manifestation of this unphysical effect can be seen very clearly in the ground-state HF wave functions of an $N\ne Z$ nucleus calculated 
by using an isospin-invariant interaction~\cite{Satula2,Rafalski,Satula1}. In the present work, we use the many-body eigenstates of the shell-model Hamiltonian, which in a spherical harmonic-oscillator
basis are characterized by the good values of angular momentum and isospin.
The isospin-mixing may only influence the radial part of the single-particle wave functions, which we use
to replace the harmonic oscillator wave functions.
Our solution to this problem will be discussed later in subsection~\ref{fit}. 

The validity of the Slater approximation~\eqref{Sl} for the Coulomb exchange potential and further electromagnetic corrections such as those due to the finite size of nucleons, vacuum polarization, electromagnetic spin-orbit as well as the impact of charge density gradient on the approximation just mentioned, will be discussed together in section~\ref{EM}. 

\subsection{Center-of-mass corrections}\label{ccom} 

As a common practice for the mean-field calculations~\cite{Vautherin}, the center-of-mass (COM) correction 
is done by subtracting the COM kinetic energy $\boldsymbol{P}^2/2mA$ from the many-particle Hamiltonian. 
As the total momentum $\boldsymbol{P}=\sum_{i=1}^A\boldsymbol{p}_i$ is a sum of single-particle momenta, the COM kinetic energy 
can be written as a sum of two terms:
\begin{equation}\label{com}
\frac{1}{2mA}\boldsymbol{P}^2 = \frac{1}{2mA} \sum_{i=1}^A \boldsymbol{p}_i^2+\frac{1}{2mA}\sum_{i\ne j} \boldsymbol{p}_i\cdot\boldsymbol{p}_j. 
\end{equation}
The effects of the first term can be included by multiplying the factor $\hbar^2/2m$ in the kinetic energy
term of Eq.~\eqref{com} by a factor $(A-1)/A$. Equivalently, the mass is increased by a factor $A/(A-1)$. The second term gives rise only to exchange corrections. However, it is difficult to calculate and is usually neglected for the calculations using the Skyrme-type interaction. 
With two-body term in Eq.~\eqref{com} being neglected, the resulting COM correction are denoted here as \emph{type A COM correction}.  

On the other hand, another philosophy has been adopted to remove the spurious COM contribution with the WS potential. In this potential model, a nucleus is considered as a two-body system comprising a core of $(A-1)$ nucleons and a nucleon outside the core. Subsequently, the COM correction can be done by simply replacing the nucleon mass with a reduced mass of $\mu=m(A-1)/A$~\cite{XaNa2018}. 
It is denoted as \emph{type B COM correction}. 

\begin{figure}[htb]
\begin{center}
\includegraphics[]{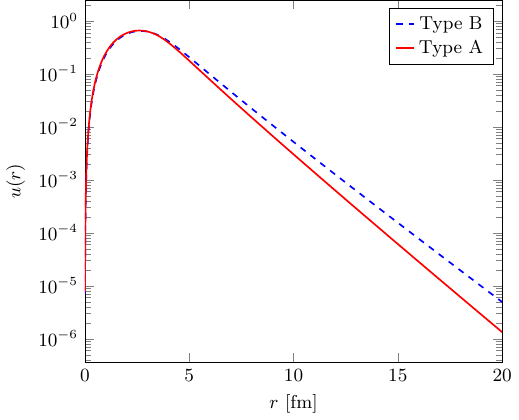}
 \caption{(Color online) Comparison between the $1p_{1/2}$ proton radial wave function obtained from a Skyrme-HF calculations using the type A and type B COM corrections (see the main text for details).}
\label{fig5}
\end{center}
\end{figure}

As corrections of both types directly affect the binding energy, the tails of WS and HF radial wave functions will be slightly different even if their eigenvalues are fitted to match the same experimental data. It can be noticed that the type B correction based on the two-body picture is more realistic at large separation distance, when the contribution from the internal structure of the $(A-1)$ core can be neglected. Since the type B correction increases the kinetic energy, the single-particle states will be less bound and more extended towards the classically forbidden region. As a result, the radial wave functions will be more sensitive to fine details of the mean field, including the Coulomb and density-driven (a-)symmetry terms. Therefore, within this correction type, we will get a larger value for $\delta_{RO}$. The effect of the type A correction is opposite. Evidently, this argument supports the difference between the Ormand-Brown's and Towner-Hardy's result for $\delta_{RO}$. However, we have numerically checked that the impact of this induced mass uncertainty on $\delta_{RO}$ is quite small ($\sim 0.012~\%$ or less), especially for heavy nuclei. Since the Skyrme-interaction parametrizations are usually fitted with the type A correction, we have no reason not to use it for the present calculations. 

\subsection{Potential adjustment and suppression of spurious isospin mixing}\label{fit}

As a basic requirement of the full parentage formalism for $\delta_{RO}$~\ref{forma}, the independent-particle potential must be adjusted in order to get the correct asymptotic form of radial wave functions. 
According to the Ormand-Brown's protocol~\cite{OrBr1985}, the local equivalent potential $\tilde{U}_{\alpha,q}^L(r)$ obtained from a Skyrme HF calculation is adjusted by scaling its central term, so that the energy eigenvalues match the experimental separation energies, whereas its repulsive Coulomb, spin-orbit, energy dependence and effective mass derivative terms are all kept fixed. 
This procedure seems analogous to that applied in a series of the shell-model calculations with WS potential of Towner and Hardy~\cite{ToHa2002,ToHa2008,HaTo2015}, except that the charge radii calculated with HF radial wave functions are not fitted to the experimental data. 

It should be noted, however, that Ormand and Brown did not correct for the spurious isospin mixing which is a byproduct of the HF approximation, even in the absence of charge-dependence forces, due to the self-consistency of the potential, $\tilde{U}_{\alpha,q}^L(r)$. 
Recently, it was found within a $JT$-projected DFT calculation~\cite{Satula2,Rafalski,Satula1} that, the contribution of this unphysical effect could be as large as 30~\% of the value of the isospin-mixing parameter. It is thus interesting to examine the impact of this spurious symmetry breaking on the calculation of $\delta_{RO}$. 

Since superallowed $0^+\to 0^+$ nuclear $\beta$ decays occur at or in the vicinity of the $N=Z$ line, we can search for an approximate method to suppress the isospin spuriosity while keeping as much as possible the physical properties of the HF single-particle wave functions.
We remind again that our many-body states are expressed as a mixing of many-body spherical configurations, obtained from the shell-model diagonalization,
and they are characterized by well-defined angular momentum and isospin quantum numbers. 
What we need is only to correct for spurious isospin mixing in individual states used to replace traditional harmonic-oscillator single-particle states.
To this end, we propose the following simplified procedure. The first step and the starting point is to find an optimal isospin-invariant HF solution for a given nucleus which we suppose to be the solution of a self-consistent HF calculation without the Coulomb force for an $N=Z$ nucleus of the same isobaric chain. Next, from the obtained solution, we perturbatively construct for the parent and daughter nuclei various densities and, subsequently, find a full local equivalent potential, $\tilde{U}_{\alpha,q}^L(r)$ including the Coulomb terms. Obviously, in this step we neglect the potential rearrangement due to the addition of the Coulomb term as well as the modification in the neutron-proton configuration when moving from the $N=Z$ nucleus to $N\ne Z$ nuclei. 
Finally, the single-particle radial wave functions used to evaluate the overlap integrals, $\Omega_{k_\alpha}^\pi$, are obtained by solving Eq.~\eqref{equiv} 
using the resulting potentials, while scaling the overall strength of its central part in order to reproduce the experimental separation energies. 
Our HF calculations were carried out using the SkM* parametrization for the Skyrme interaction~\cite{skm*}. 
The spread of $\delta_{RO}$ due to the different parameterizations of the Skyrme functional was found to be small~\cite{Xthesis}. 

\begin{figure}[htb]
\begin{center}
\includegraphics[]{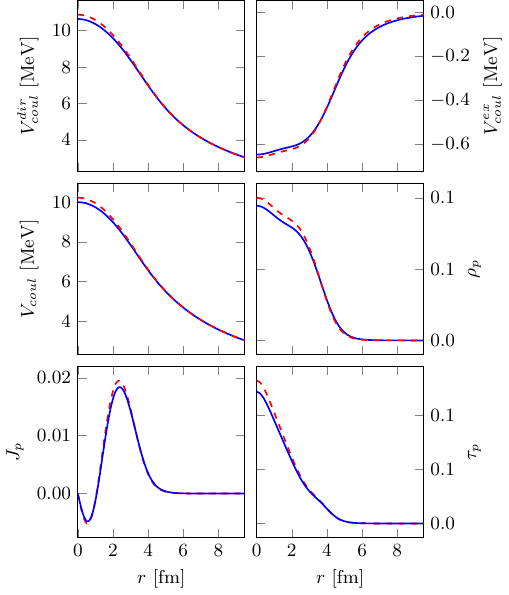}
\caption{\label{fig3}(Color online) The proton densities and the Coulomb potentials (in MeV). The solid curves represent the result of a Standard HF calculation, while the dashed curves show the same quantities after suppression of spurious isospin mixing for comparison.}
\end{center}
\end{figure}

The suppression of spurious isospin mixing gives rise to compression of proton densities. 
A comparative analysis of Fig.~\ref{fig3} shows that the magnitude of proton densities in the nucleus interior becomes larger after the suppression (except for $J_p(r)$ because of its oscillations) and drops faster or is less extended towards the asymptotic region. 
Likewise, the magnitude of the Coulomb direct term constructed from these spuriosity-free proton densities is increased in the nucleus interior, but its tail is rather insensitive against such a change in the proton densities. The Coulomb exchange term within the Slater approximation is modified in a similar manner as the proton densities, but this modification is very small in magnitude and does not seem to produce any significant contribution to $\delta_{RO}$. In contrast, neutron densities 
including $\rho_n$, $J_n$ and $\tau_n$ are almost unaffected by this isospin spuriosity suppression. 
It can thus be concluded from this result that spurious isospin mixing does not impact the asymptotic behavior of the HF radial wave functions. 

In general a more repulsive Coulomb potential would result in a larger value of $\delta_{RO}$. 
However, this is only true for some transitions and not for the others as one can see from Fig.~\ref{fig8} (the corresponding numerical values are listed in Table~\ref{tab3}). 
The reason is that the change in the radial form of the proton densities affects not only the Coulomb potential but also the nuclear part such as the central and spin-orbit terms. This gives rise to a modification of the isovector property of the Skyrme HF nuclear field which, in turn, softens the effect of the Coulomb repulsion on $\delta_{RO}$. We recall that the Coulomb strength increases with the atomic number, whereas the strength of the isovector potential is approximately proportional to the difference between the neutron and proton densities ($\propto\rho_n-\rho_p$)~\cite{DG} which, for our cases, is nearly constant since we study superallowed $\beta$ decay of nuclei near the $N=Z$ line. Therefore, the Coulomb effect should be dominated in medium to heavy mass emitters while the isovector effect could only be dominated over the Coulomb one in very light mass emitters. In addition, the potential adjustment also slightly modifies the isovector property whose net effect is determined by the difference between the calculated and experimental $CDE$ values. 

\section{Electromagnetic corrections}\label{EM} 
\subsection{Slater approximation}\label{slat}

\begin{figure}[htb]
\begin{center}
\includegraphics[]{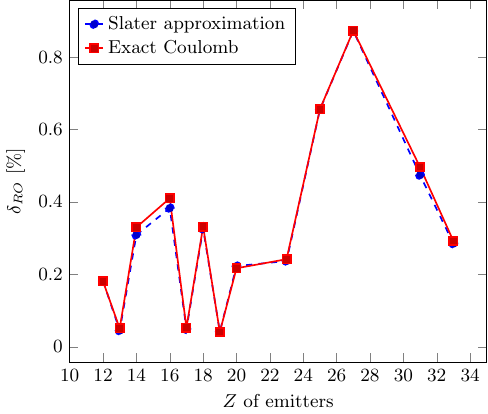}
\caption{(Color online) $\delta_{RO}$ values for some selected transitions, obtained with the Slater approximation and the exact treatment of the Coulomb exchange term. }
\label{fig6}
\end{center}
\end{figure}

The description of the structural properties of atomic nuclei based on the Slater approximation for the Coulomb exchange term 
has been already discussed by various authors~\cite{Skalski,LeBloas,Titin}. For spherical closed sub-shell nuclei, the relative error associated with the use of this approximation decreases with the nucleon number from about 8~\% in $^{16}$O to about 2~\% in heavy nuclei. 
It was also pointed out~\cite{HaTo2009} that the Slater approximation leads to an overestimated Coulomb potential at large distance which is the region of great interest for the present study, i.e., the sum of Eq.~\eqref{Dir} and Eq.~\eqref{Sl} falls off as $Ze^2/r$ instead of $(Z-1)e^2/r$. 
In order to check the accuracy of this approximation and its impact on the Fermi matrix element, 
we have carried out a comparative study of $\delta_{RO}$ using the code Finres4 from K.~Bennaceur~\cite{karim} which treats the Coulomb exchange term exactly. 
Note, however, that an exact HF calculation using a nonzero-range force, such as the Coulomb interaction, leads to a singular local equivalent potential~\cite{VV}, which is technically difficult to manipulate within the Ormand-Brown's protocol and the framework of the full parentage formalism~\ref{forma}. For simplicity, our calculations were performed in the closure approximation~\cite{Xthesis} and without any adjustment of the mean field potential. The results for selected transitions are shown in Fig.~\ref{fig6}. 

As shown in Fig.~\ref{fig6}, the differences between the $\delta_{RO}$ values obtained with the exact treatment and Slater approximation for the Coulomb exchange term are quite small, the maximum difference in the correction's magnitude is only about 0.027~\%. 
Furthermore, we found among all the transitions the maximum difference in the calculated $CDE$ of only about 226~keV. 
According to our sensitivity study in Ref.~\cite{Xthesis}, a $CDE$ deviation of 226~keV will cause a change of about 0.05~\% in the $\delta_{RO}$'s magnitude. Thus, it can be estimated that the errors due to the use of the Slater approximation would not exceed 0.027~\%+0.05~\%=0.077~\% (or $\sim 20~\%$ in relative error). Evidently, this minor deficiency cannot explain the strong discrepancy between the Ormand-Brown's and the Towner-Hardy's result. Although our calculation in this section is based on the closure approximation, the deficiency of the Slater approximation would not be significantly amplified when $\delta_{RO}$ is calculated in the full parentage formalism, since our adjustment procedure does not lead to a modification of the Coulomb potential. An alternative approach for treating the Coulomb exchange term is discussed in the next subsection. 

\subsection{Generalized gradient approximation}\label{gen}

Recently it has been shown that Coulomb energy density functionals built by using the generalized gradient approximation (GGA) give almost the same accuracy for the total energy as the exact treatment~\cite{Naito1,Naito1}, while its numerical cost is still the same as that of the Slater approximation. 
The Coulomb exchange potential derived with the GGA depends not only on the charge density but also on its gradient, namely, 
\begin{equation}\label{gga}
\begin{array}{ll}
\displaystyle V_{Coul}^{ex}(r) =& \displaystyle -e^2\left[\frac{3}{\pi}\rho_{ch}(r)\right]^\frac{1}{3} \left\{ F(s)-\left[s+\frac{3}{4k_F r}\right] F'(s) \right. \\[0.18in]
&+\displaystyle\left.\left[s^2-\frac{3\rho_{ch}''(r)}{8\rho_{ch}(r)k_F^2}\right]F''(s) \right\}, 
\end{array}
\end{equation}
where $\rho_{ch}(r)$ is the density of charge (see subsection~\ref{cdens} for details), $-e^2[(3/\pi)\rho_{ch}(r)]^{1/3}$ corresponds to the Coulomb exchange term obtained with the Slater approximation, $F(s)$ is the enhancement factor due to the density gradient which, using the Perdew-Burke-Ernzerhof parametrization~\cite{Perdew}, is expressed as 
\begin{equation}
\displaystyle F(s) = 1+\kappa - \frac{ \kappa }{ (1+\mu s^2/\kappa) }, 
\end{equation}
with the function $s$ denoting the dimensionless density gradient, 
\begin{equation}\label{s}
\displaystyle s = \frac{ |\boldsymbol{\nabla} \rho_{ch}(r)| }{ 2k_F\rho_{ch}(r) }, 
\end{equation}
where the local Fermi wave number is given by $k_F=[3\pi^2\rho_{ch}(r)]^{1/3}$. 
The parameters $\kappa=0.804$ and $\mu=0.274$ are taken from Ref.~\cite{Naito1}. 

Note that in order to shorten the length of Eq.~\eqref{gga}, we have written $F'(s)$ and $F''(s)$, respectively, for the first and second derivatives of the enhancement factor $F$ with respect to the function $s$, and $\rho_{ch}''(r)$ for the second derivative of charge density with respect to the radial variable $r$. This is inconsistent with the notation used in the previous subsection. 

Obviously, if the nuclear charge distribution is slowly varying with the distance $r$, the GGA would reduce to the Slater approximation. 
However, in reality, the gradient $|\boldsymbol{\nabla} \rho_{ch}(r)|$ always has a finite value, especially in the surface region where the charge density drops very quickly. It is shown in the top-right panel of Fig.~\ref{fig7} that the Coulomb exchange potential derived within the GGA diverges significantly from that derived within the Slater approximation. 
In particular, it becomes more oscillating and extended to large distances when the charge density gradient is taken into account. 
It should be noticed also that the GGA does not restore the asymptotics of the exact Coulomb potential of $(Z-1)e^2/r$. 

In general, the correction values obtained with the GGA are significantly larger than those obtained with the Slater approximation. More detail will be discussed in section~\ref{result}. 

\subsection{Vacuum polarization correction}

Another correction to the Coulomb interaction between protons is the vacuum polarization, that is the
effect of quantum fluctuations in the vacuum. The direct electromagnetic interaction happens just through the exchange of a photon between the charged particles. 
However, an electromagnetic field will, in general, induce a charge and current distribution due to the creation and annihilation of virtual particle-antiparticle pairs. 
The contribution of vacuum polarization to the Coulomb potential is given by~\cite{Uehling1935,FulRin1976}, 
\begin{equation}\label{vp}
\begin{array}{ll}
\displaystyle V_{VP}(r) =& \displaystyle \frac{2\alpha e^2\lambdabar_e}{3r}\int_0^\infty dr'r'\rho_{ch}(r) \left[ K_0\left(\frac{2}{\lambdabar_e}|r-r'|\right) \right. \\[0.18in]
& \displaystyle  \left.-K_0\left(\frac{2}{\lambdabar_e}|r+r'|\right) \right], 
\end{array}
\end{equation}
where $\lambdabar_e={\hbar}/{m_ec}=86.159$~fm is the reduced electron Compton wavelength, and $\alpha=1/137.036$ is the fine-structure constant. We remark that the vacuum polarization potential is always positive, so it gives a repulsive contribution to the interaction. Here we neglect the higher order
terms that are smaller, since they are proportional to
higher powers of the fine-structure constant~\cite{Uehling1935,FulRin1976}. The function $K_0$ is defined as,  
\begin{equation}\label{ko}
K_0(x) = \int_1^{+\infty}dt\left[e^{-xt}\left(\frac{1}{t^2}+\frac{1}{2t^5} \right)\sqrt{t^2-1}\right], 
\end{equation}
where $x$ and $t$ are dimensionless variables. We notice that the computation of the function $K_0$ using Eq.~\eqref{ko} is very time-consuming because the numerical integration over the variable $t$ is very slowly convergent. To avoid the numerical difficulty, we approximate the function $K_0$ with the following fifth order polynomial which was obtained from a least-square fitting:  
\begin{equation}\label{poly}
\begin{array}{lll}
K_0(x) &=& -5575.506x^5+2629.071x^4-491.189x^3 \\[0.18in]
     && +50.496x^2-4.804x+0.874, 
\end{array}
\end{equation}
we have numerically checked the validity of this approximation over the domain of interest. 

\begin{figure}[htb]
\begin{center}
\includegraphics[]{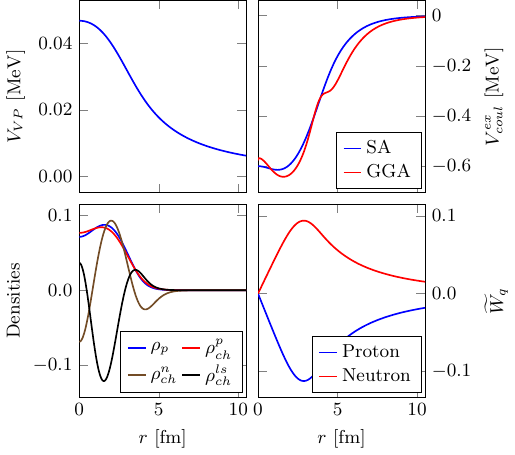}
 \caption{(Color online) The repulsive Coulomb potential due to the vacuum polarization (top-left), the Coulomb exchange potential obtained with the GGA and the Slater approximation (SA) (top-right), the comparison between the proton density and charge densities (bottom-left) where $\rho_{ch}^n$ and $\rho_{ch}^{ls}$ are scaled by multiplying with 100, and the Coulomb spin-orbit potential (bottom-right). The unit for the potentials is MeV.}
\label{fig7}
\end{center}
\end{figure}

Although, the potential $V_{VP}(r)$ is repulsive and has a long tail as one can see from the top-left panel of Fig.~\ref{fig7}, it is two orders of magnitude smaller than the Coulomb direct potential. We found that the inclusion of $V_{VP}(r)$ results in a smaller value of $\delta_{RO}$. 
Nevertheless, the vacuum polarization leads to a decrease of $\delta_{RO}$'s by $1$~\% -- $3$~\% (in relative magnitude). 
Therefore, this effect is completely negligible comparing to the uncertainties from the effective shell-model interactions and 
the Skyrme-force parametrizations~\cite{Xthesis}.  

\subsection{Electromagnetic spin-orbit correction}\label{coulso}

In addition to the Coulomb potentials identified in the previous subsection, the relativistic correction such as the Coulomb spin-orbit effect could be important for a nucleon with high Fermi momentum. 
This correction can be expressed in the famous Thomas-form~\cite{Thomas} as follows: 
\begin{equation}\label{cso}
V_{cso}^q(r) = \frac{1}{r} \widetilde{W}_q(r) \braket{\boldsymbol{\sigma}\cdot\boldsymbol{l}}, 
\end{equation}
the radial form factor, $\widetilde{W}_q(r)$ is given by
\begin{equation}\label{cform}
\widetilde{W}_q(r) = \frac{1}{4}\left(\frac{\hbar}{mc}\right)^2g_q'\frac{d}{dr}V_{Coul}(r), 
\end{equation}
where $g_q'$ is equal to $g_p-1$ for protons and $g_n$ for neutrons; $g_n=-3.82608545(90)$ and $g_p=5.585694702(17)$ are the
neutron and proton gyromagnetic factors, respectively~\cite{Mohr}. 

As with the nuclear spin-orbit term which is responsible for the shell structure in nuclei, the Coulomb spin-orbit potential~\eqref{cso} is also proportional to the spin-orbit coupling and the gradient norm of the associated mean field, except that its strength is smaller than that of the nuclear one by almost two orders of magnitude. Furthermore, the first order derivative in Eq.~\eqref{cform} is always negative since the total Coulomb potential is a monotonically-decreasing function of $r$. Therefore, the form factor $\widetilde{W}_q(r)$ is negative for protons and positive for neutrons as can be seen from the bottom-right panel of Fig.~\ref{fig7}, whereas the nuclear spin-orbit form factor for a given nucleon type can oscillate around the $r$ axis, in particular, in the nucleus interior. 

Writing the expectation value $\braket{\boldsymbol{\sigma}\cdot\boldsymbol{l}}$ explicitly in the following form 
\begin{equation}
\braket{\boldsymbol{\sigma}\cdot\boldsymbol{l}} = \left\{
\begin{array}{ll}
l,      &  \text{if $j=l+1/2$} \\[0.18in]
-(l+1), &  \text{if $j=l-1/2$}, 
\end{array}
\right.
\end{equation}
it is evident that the Coulomb spin-orbit potential for protons is attractive for spin-up states and repulsive for spin-down states, while its effect for neutrons is opposite since the neutron $g$-factor is negative. The contribution of this relativistic correction to $\delta_{RO}$ is of the same order of magnitude as that of the vacuum polarization. However, it greatly varies from one transition to another within the range between $-2~\%$ and $3.7~\%$ of $\delta _{RO}$. 

\subsection{Finite nucleon size correction}\label{cdens}

All the Coulomb terms discussed above are, in principle, written in terms of the nuclear charge density $\rho_{ch}(r)$, because the Coulomb interaction affects the charge itself instead of the pointed protons. Basically, the nuclear charge density can be written as a sum of three terms~\cite{Bertozzi,Friar,Chandra}, $\rho_{ch}^p(r)$ which comes from the finite charge distribution of the proton folded with the pointed proton density, $\rho_{ch}^n(r)$ which comes from the finite charge distribution of the neutron folded with the pointed neutron density, and $\rho_{ch}^{ls}(r)$ which is a relativistic electromagnetic correction which depends on the spin-orbit coupling $\boldsymbol{l}\cdot\boldsymbol{\sigma}$. Thus we have
\begin{equation}\label{oo}
\rho_{ch}(r) = \rho_{ch}^p(r) + \rho_{ch}^n(r) + \rho_{ch}^{ls}(r), 
\end{equation}
where the first and second terms are by definition given by
\begin{equation}\label{ch}
\rho_{ch}^q(\boldsymbol{r}) = \int d\boldsymbol{r'} \rho_q(\boldsymbol{r'})G_q(\boldsymbol{r}-\boldsymbol{r'}). 
\end{equation}

The effective electromagnetic form factor $G_q$ is taken as the sum of three Gaussians for proton ($n_p=3$) and two Gaussians for neutron ($n_n=2$). They are collectively expressed as
\begin{equation}\label{for}
\displaystyle G_q(\boldsymbol{r}) = \sum_{i=1}^{n_q} \frac{a_q^i}{(r_q^i\sqrt{\pi})^3} \exp{\left[-\frac{\boldsymbol{r}^2}{(r_q^i)^2}\right]}, 
\end{equation} 
where the parameters $a_q^i$ and $r_q^i$ are taken from Ref.~\cite{Chandra}. 
It can be noticed that the proton charge radius of 0.88~fm deduced from this sum of three Gaussians is a little larger than the value of 0.83~fm~\cite{Pohl} needed if a dipole form factor is used. However, as pointed out by Chandra and Sauer~\cite{Chandra}, the proton charge form factor for high momentum is not well fitted by the dipole form. 

By performing the analytical integration over the angular part of Eq.~\eqref{ch}, we get
\begin{equation}
\begin{array}{ll}
\displaystyle \rho_{ch}^q(r) = & \displaystyle \sum_i^{n_q} \frac{a_q^i}{\sqrt{\pi}}\int_0^\infty dr' \frac{r'\rho_q(r')}{r_q^ir}\left\{ \exp\left[-\left( \frac{r-r'}{r_q^i} \right)^2 \right] \right. \\[0.18in]
& \displaystyle \left. -\exp\left[-\left( \frac{r+r'}{r_q^i} \right)^2 \right] \right\}.
\end{array}
\end{equation}

The correction due to spurious center-of-mass motion and the Darwin-Foldy
relativistic correction to the charge density can be included in Eq.~\eqref{for} by replacing $(r_q^i)^2$ by
\begin{equation}\label{corr}
\displaystyle (r_q^i)^2 + \frac{1}{2}\left(\frac{\hbar}{mc}\right)^2-\frac{b^2}{A}, 
\end{equation}
where the oscillator parameter is taken as $b^2=41.465/\hbar\omega\approx A^{1/3}$~fm$^2$ and $\hbar/mc=0.21$~fm. 

The relativistic spin-orbit contribution to charge density using the factorization approximation introduced by Bertozzi {\sl et al.}~\cite{Bertozzi} is given by
\begin{equation}\label{zzz}
\displaystyle\rho_{ch}^{ls}(r) =-\left(\frac{\hbar}{mc}\right)^2 \sum_{\alpha,q} \nu_\alpha^q\braket{\boldsymbol{\sigma}\cdot\boldsymbol{l}}g_q'\frac{1}{r^2}\frac{d}{dr}\Big[r\rho_\alpha^q(r)\Big] , 
\end{equation} 
where $\rho_\alpha^q(r)$ and $\nu_\alpha^q$ are the point-particle density and the occupation number of the orbit $\alpha$, respectively. The constants $g_q'$ are given in section~\ref{coulso}. Note that this spin-orbit contribution is only important for the orbits which have $\nu_\alpha^p\ne \nu_\alpha^n$, thus it should be small for nuclei with $N\approx Z$. 

The contribution $\rho_{ch}^p(r)$ is generally more extended and dilute than the point-proton density, therefore the finite size effect of protons makes the Coulomb interaction effectively weaker. 
We notice that the deviation of the total charge density~\eqref{oo} from the point-proton density is dominated by $\rho_{ch}^p(r)$. The contribution due to the finite size of neutrons, $\rho_{ch}^n(r)$ cancels out almost completely with the relativistic spin-orbit contribution, $\rho_{ch}^{ls}(r)$. The comparison is displayed in the bottom-left panel of Fig.~\ref{fig7}. 

The effect of finite nucleon size on $\delta_{RO}$ varies with the mass number, but in the range between $-0.4~\%$ (in heavy nuclei) and $3~\%$ (in light nuclei) only. However, the inclusion of finite size correction leads to a significant increase of charge radii which, according to the Towner-Hardy's protocol~\cite{ToHa2002,ToHa2008}, can be viewed as an indirect effect on $\delta_{RO}$. More details are discussed in subsection~\ref{radii}. 

\section{Charge-symmetry-breaking and charge-independence-breaking forces}\label{isb}

We distinguish here two kinds of isospin-symmetry-breaking nuclear interactions which are known to be important to solve the Nolen-Schiffer anomaly problem~\cite{Nolen}: the one that breaks charge independence (CIB) and the one that breaks charge symmetry (CSB). A CIB force means that $v_{np}\ne(v_{pp}+v_{nn})/2$, while a CSB force means that $v_{pp}\ne v_{nn}$. In fact, experiments show that $v_{nn}$ is about 1~\% stronger than $v_{pp}$, and $v_{np}$ is about 2.5~\% stronger than the average of $v_{nn}$ and $v_{pp}$~\cite{Machleidt}. The effects of isospin non-conserving interactions of nuclear origin on the superallowed $0^+\to0^+$ nuclear $\beta$ decays have been discussed in the shell-model calculations of the isospin-mixing component, $\delta_{IM}$~\cite{OrBr1985,HaTo2015}, but not in the calculations of $\delta_{RO}$. We take both of them into account in our HF calculations by adding to the conventional Skyrme interaction the following CIB and CSB terms~\cite{Sagawa}: 
\begin{equation}\label{CIB}
\begin{array}{lll}
v_{CIB} &=&\displaystyle \frac{1}{2} t_{CIB} \delta \left[ P_0^{uz} + \frac{1}{2}P_1^{uz}\left(\boldsymbol{k}^2+\boldsymbol{k}'^2\right) + P_2^{uz}\boldsymbol{k}'\cdot\boldsymbol{k} \right], \\[0.18in]
v_{CSB} &=&\displaystyle \frac{1}{2} t_{CSB} \delta \left[ P_0^{sy} + \frac{1}{2}P_1^{sy}\left(\boldsymbol{k}^2+\boldsymbol{k}'^2\right) + P_2^{sy}\boldsymbol{k}'\cdot\boldsymbol{k} \right],
\end{array}
\end{equation}
where $t_{CIB}=4t_{iz}t_{jz}$ and $t_{CSB}=2(t_{iz}+t_{jz})$. The operators $P_i^{uz}$ and $P_i^{sy}$ are related to the spin-exchange operator as $P_i^{sy}=s_i(1+y_iP_\sigma)$ and $P_i^{uz}=u_i(1+z_iP_\sigma)$. The $u_i$ and $s_i$ are free parameters which one can find with a fit to experimental data. For our calculations, we use the the parameters' value extracted from the strength and the range of the Yukawa interaction as discussed in Ref.~\cite{Sagawa}. All exchange parameters are $y_i=z_i=-1$ because of the singlet-even character of the isospin-breaking forces. 

Note that, in principle, a CIB force should have an isotensor structure, namely, $\propto (t_{iz}t_{jz}-\boldsymbol{t}_{i}\cdot\boldsymbol{t}_{j}/3)$ as described in Ref.~\cite{Baczyk2018}. For the present work, we simply neglect the term, $\boldsymbol{t}_{i}\cdot \boldsymbol{t}_{j}/3$ because it contributes a neutron-proton mixing component to the mean field potential. 

The HF energy density functional associated to the CSB and CIB interactions~\eqref{CIB} is, respectively, 
\begin{equation} 
\begin{array}{ll}
\displaystyle\mathcal{H}_{CSB}(r) 
&=\displaystyle\frac{1}{16}\sum_q t_{qz}\Big\{ 4s_0(1-y_0)\rho_q^2(r)+s_1(1-y_1) \\[0.18in]
&\times\displaystyle\left[ -\frac{3}{2}\rho_q(r)\Delta\rho_q(r)+J_q^2(r)+2\rho_q(r)\tau_q(r) \right] \\[0.18in]
&+\displaystyle \left[ \frac{3}{2}\rho_q(r)\Delta\rho_q(r)-J_q^2(r)+6\rho_q(r)\tau_q(r) \right] \\[0.18in]
&\times\displaystyle s_2(1+y_2) \Big\}, 
\end{array}
\end{equation}
and 
\begin{equation} 
\begin{array}{ll}
\mathcal{H}_{CIB}(r) 
&=\displaystyle \frac{1}{32}\sum_q\Big\{ 4\left[ u_0(1-z_0)\rho_q^2(r)-u_0(2+z_0) \right. \\[0.18in]
&\times\displaystyle \left.\rho_q(r)\rho_{q'}(r) \right]+\frac{3}{2}\rho_q(r)\left[ u_1(2+z_1)\Delta\rho_{q'}(r) \right. \\[0.18in]
&-\displaystyle \left. u_1(1-z_1)\Delta\rho_q(r) \right]+2u_1(1-z_1)\rho_q(r)\tau_q(r) \\[0.18in]
&-\displaystyle 2u_1(2+z_1)\rho_q(r)\tau_{q'}(r)+u_1(1-z_1)J_q(r) \\[0.18in]
&+\displaystyle u_1z_1J_q(r)J_{q'}(r) \Big\}. 
\end{array}
\end{equation}

Due to their momentum-dependence components, both the CIB and CSB forces~\eqref{CIB} contribute not only to the central part of the HF mean field, but they also induce a perturbation to the structure of the kinetic energy and spin-orbit potentials. The contribution of the CIB and CSB forces to the effective mass is derived to be 
\begin{equation}
\begin{array}{lll}
\displaystyle \frac{\hbar^2}{2m_q^*(r)} = \frac{1}{16} \Big[ 2(u_1 + 2t_{qz}s_1)\rho_q(r) + \left( u_1+u_2\right)\rho_{q'}(r) \Big], 
\end{array} 
\end{equation}
the contribution to the central potential is  
\begin{equation}\label{e2.33}
\begin{array}{lll}
U_q(r) 
&=& \displaystyle\frac{u_0}{4}\left[2\rho_q(r)-\rho_{q'}(r)\right]+ t_{qz}s_0\rho_q(r) \\[0.18in]
&-& \displaystyle\frac{3}{16}u_1\Delta \rho_q(r) +\frac{1}{32}\left[3u_1 -u_2\right]\Delta \rho_q(r) \\[0.18in]
&+& \displaystyle\frac{1}{8}u_1\tau_{q}(r)-\frac{1}{16}\left[u_1+u_2\right]\tau_{q'}(r) \\[0.18in]
&-& \displaystyle \frac{1}{2}t_{qz}\left[ \frac{3}{4}s_1\Delta \rho_q(r)+s_1\tau_q(r) \right], 
\end{array}
\end{equation}
and to the spin-orbit potential is 
\begin{equation}\label{e2.34}
\displaystyle W_q(r)=\frac{1}{8}\left(u_1+2t_{qz}s_1\right)J_q(r) - \frac{1}{16}\left(u_1+u_2\right)J_{q'}(r).
\end{equation} 

Although these contributions are small in magnitude, they are strongly charge-dependent and could potentially cause a significant modification to the isovector property of the Skyrme HF mean field. The effects of these extended forces on $\delta_{RO}$ will be discussed in section~\ref{result}.

\section{Results and discussions}\label{result} 
\subsection{Radial overlap correction}

\begin{figure*}[htb]
\begin{center}
\includegraphics[]{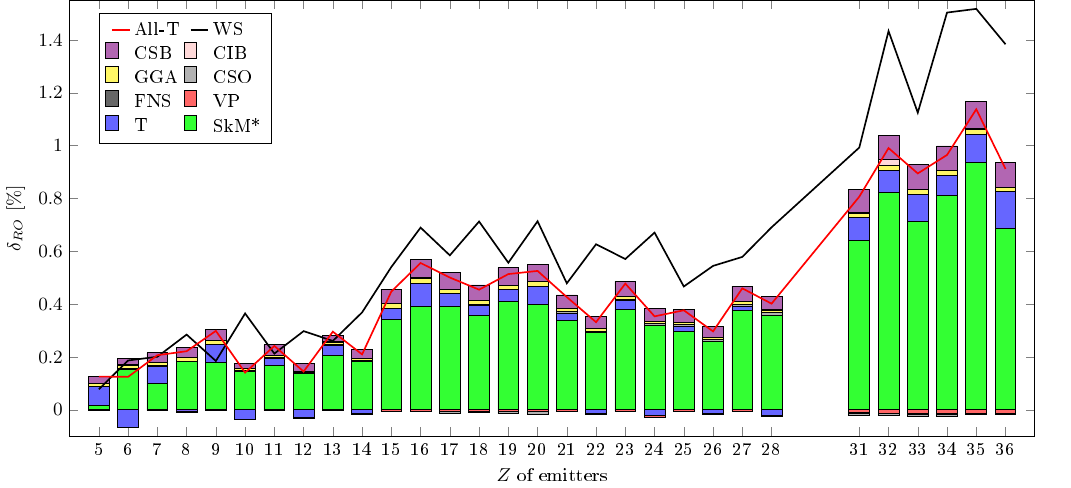}
\caption{\label{fig8}(Color online) Results for the radial overlap correction. The values denoted with "WS" are obtained WS radial wave functions. The others are obtained with HF radial wave functions using the Skyrme interaction~\cite{skm*} and the extensions as detailed in the main text. For a better visualization, we plot T, VP, FNS, CSO, GGA, CIB and CSB as the individual contribution due to the suppression of spurious isospin mixing, vacuum polarization, finite nucleon size, Coulomb spin-orbit, effect of charge density gradient, charge-independence-breaking and charge-symmetry-breaking forces, respectively. In the other word, the contribution T is obtained by subtracting SkM* from SkM*-T. Whereas the contribution VP, FNS, CSO, GGA, CIB and CSB are obtained by subtracting SkM*-T from VP-T, FNS-T, CSO-T, GGA-T, CIB-T and CSB-T, respectively.}
\end{center}
\end{figure*}

Nine different calculations of $\delta_{RO}$ based on the Skyrme-HF radial wave functions have been carried out for each transition under consideration. The results are displayed in Fig.~\ref{fig8} and also listed in Table~\ref{tab3}. 
It turns out that the contribution of the electromagnetic corrections such as the vacuum polarization (VP), finite nucleon size (FNS) and Coulomb spin-orbit (CSO), as well as that of the CIB force are not significant. 
We recall from section~\ref{EM} that the contribution of the vacuum polarization is negative but does not exceed $-3$~\% (in relative percentage), the contribution of the Coulomb spin-orbit varies between $-2~\%$ and 3.7~\% depending strongly on the single-particle configuration, and that of the finite nucleon size, which is dominated by the internal charge distribution of protons, has a similar order of magnitude, except that it is positive for light mass emitters and becomes negative for heavier emitters, whereas the contribution of the CIB force is even smaller. 

Among the other five correction sets, the one named ``SkM*'' was obtained with the Ormand-Brown's protocol~\cite{OrBr1985} employing a conventional Skyrme interaction which, apart from the Coulomb terms, is invariant under rotations in isospin space~\cite{skm*}. 
This result is, however, significantly different from the values reported in Ref.~\cite{OrBr1995}. 
The principal reasons are that larger configuration spaces and more recent effective shell-model interactions were implemented in the present work. 
The values labeled ``SkM*-T'' are also obtained with SkM*, while the additional label "-T" indicates that this correction set was obtained 
with the suppression of spurious isospin mixing as discussed in subsection~\ref{fit}. 
The sets labeled ``GGA-T'' and ``CSB-T'' were obtained by using the GGA instead of the Slater approximation for the Coulomb exchange term, and by adding the CSB force within the conventional effective Skyrme interaction, respectively. Other details are the same as the "SkM*-T" calculation. 
While for the set labeled ``All-T'', all the extensions and developments were simultaneously taken into account. 
For comparison, the result obtained on equal footing but with WS radial wave functions instead of HF ones, 
is also plotted in Fig.~\ref{fig8} as ``WS''. For the latter calculation, we used the BM$_m$ parametrization of the WS potential 
from Ref.~\cite{XaNa2018}.

In general, the "SkM*" values are significantly lower than those obtained using the phenomenological WS potential. This discrepancy has been known since the pioneering work of Ormand and Brown~\cite{OrBr1985,OrBr1995}. 
It is, however, interesting to point out here that the global trend of both sets of correction values as a function of the atomic number of emitters are quite similar. It is seen that the correction is rapidly increased from $Z=13$ to $16$ and $Z=25$ to $31$, while it is nearly constant for the other regions, except for the local variation effects. 
We notice that the increase of $\delta_{RO}$ from $Z=13$ to $16$ is dominated by the contribution of the $2s_{1/2}$ orbital, for which the centrifugal barrier is not present. For the increase from $Z=25$ to $31$, the reason is probably the contribution of the $2p_{1/2}$ and $2p_{3/2}$ orbitals which are very weakly bounded for protons, and moreover, their centrifugal barrier is not so high comparing with that of the nearby $1f_{7/2}$ and $1f_{5/2}$ orbitals. Both calculations are in remarkable agreement for the light mass region ($A\le 26$), except that their local odd-even staggering effects appear in opposite phase. 

Because of the competition between the Coulomb and nuclear isovector potentials as discussed in subsection~\ref{fit}, the suppression of spurious isospin mixing gives different effects on $\delta_{RO}$ in different nuclei. It is seen from Fig.~\ref{fig8} that the "SkM*-T" calculation yields, for most cases, a larger value for $\delta_{RO}$ relative to those obtained with the Ormand-Brown's protocol, especially for the light odd-odd $N=Z$ emitters. 
There are, however, nine cases for which "SkM*-T" yields smaller $\delta_{RO}$ values, namely, $^{10}$C, $^{14}$O, $^{18}$Ne, $^{22}$Mg, $^{26}$Si, $^{42}$Ti, $^{46}$Cr, $^{50}$Fe and $^{54}$Ni. 
We remark that all of these emitters are even-even $N\ne Z$ nuclei. 
We found that the relative deviation of $\delta_{RO}$ between the "SkM*-T" and "SkM*" sets varies between $-42$~\% and 37~\%. 
An exclusion to this range is $^{10}$B for which the "SkM*" calculation gives very small correction value (see Table~\ref{tab3}), the resulting relative deviation is subsequently very large, namely 418~\%. We remark also that this spurious isospin suppression does not improve the local variation of $\delta_{RO}$--this behavior is still in disagreement with the "WS" calculation.  

It is also seen from Fig.~\ref{fig8} that, a further increase of 2-14~\% in the correction values can be obtained by treating the Coulomb exchange term within the GGA instead of the Slater approximation. It is interesting to note that this density gradient effect trends to be stronger for light emitters, namely about $14~\%$ in $^{10}$B and only about 2~\% in $^{70}$Kr. This result is in good agreement with the simple estimation we have made in subsection~\ref{slat}. Similarly, the correction can be increased more by 10-30~\% with the addition of the CSB force~\eqref{CIB}. We notice that this CSB effect is also consistent with the recent DFT calculations carried out by Konieczka \emph{et al.}~\cite{Konieczka2021} and the calculations using the \emph{complex} excited VAMPIR approach carried out by Petrovici \emph{et al.}~\cite{Petrovici2008}. However, our final result, "All-T" is still significantly lower than "WS", except for light emitters with $A\le26$. 

Furthermore, there appears to be an odd-even staggering on the correction set calculated with WS radial wave functions, namely, 
for a given pair of mirror transitions (fixed mass number), the $\delta_{RO}$ value obtained for an even-even emitter is generally considerably larger than that obtained 
for an odd-odd neighbor. 
We found that this effect is mainly originated due to a similar staggering of the experimental Coulomb displacement energies, shown in Fig.~\ref{figx}, 
through the potential adjustment. In contrast, this property is not well reproduced by the various calculations using HF radial wave functions, 
except for the six emitters in the $2p1f_{5/2}1g_{9/2}$ model space~\cite{jun45}. 
In fact, a similar staggering is also inherent to the experimental $ft$ values~\cite{HaTo2015}, even through the uncertainties on the data for even-even emitters are quite large. Thus, from the inspection of Eq.~\eqref{Ft}, we immediately conclude that the WS result would do a better job in bringing these data into a nucleus-independent $\mathcal{F}t$ value. 
In our opinion, the problem of the HF results for the above-mentioned cases might be due to correlations and deformation effects contained in the data that we have used for constraining the potential depth. Since the correlation energy is not measurable it is difficult to remove it properly from the measured separation energies.

\subsection{Charge radii}\label{radii}

\begin{figure*}[htb]
\begin{center}
\includegraphics[]{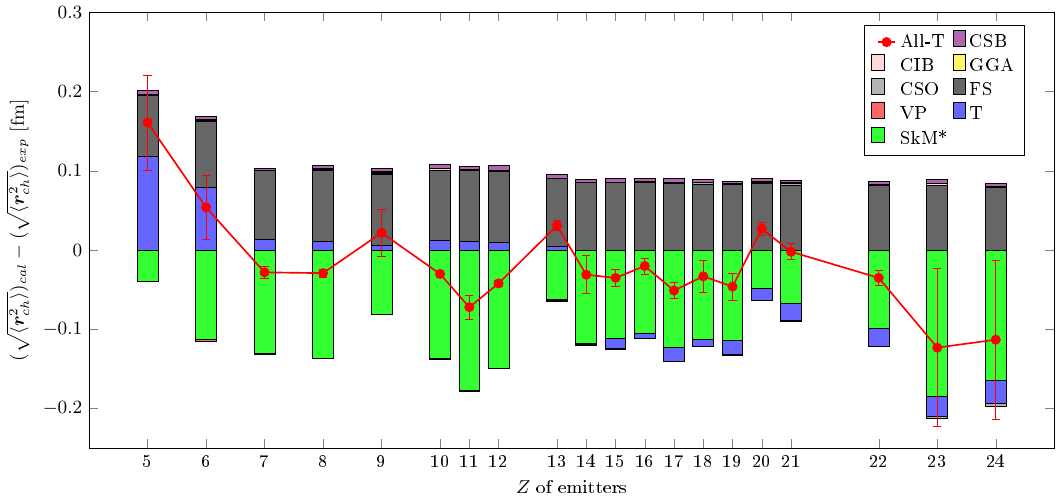}
\caption{\label{fig9}(Color online) The deviation of the calculated charge radii from the experimental data~\cite{AngMari2013,ToHa2002} in function of the atomic number of emitters. Similar to Fig.~\ref{fig8}, T, VP, FNS, CSO, GGA, CIB and CSB represent the individual contribution due to the suppression of spurious isospin mixing, vacuum polarization, finite nucleon size, Coulomb spin-orbit, effect of charge density gradient, charge-independence-breaking and charge-symmetry-breaking forces, respectively.}
\end{center}
\end{figure*}

We have also calculated the charge radius of the parent nuclei using the radial wave functions obtained from the different Skyrme-HF calculations as previously applied in the calculations of $\delta_{RO}$. The deviations of the theoretical values from the experimental data~\cite{AngMari2013,ToHa2002,Bao} are plotted in Fig.~\ref{fig9}. Note that, unlike the usual approach, our calculations employed the method proposed in Ref.~\cite{XaNa2018} with which the expectation value of the squared radius operator is expressed in the full parentage formalism, namely
\begin{equation}\label{chrad}
\begin{array}{ll}
\displaystyle \braket{\boldsymbol{r}_{ch}^2} = & \displaystyle \frac{1}{Z}\sum_{k_\alpha\pi} \left(C^p_{T_\pi}\right)^2 S_{k_\alpha,\pi}^i \int_0^\infty r^4 |R_{k_\alpha,p}^\pi(r)|^2dr \\[0.18in] 
& \displaystyle + \braket{\boldsymbol{r}_p^2} + \braket{\boldsymbol{r}_n^2} + \braket{\boldsymbol{r}_{ls}^2} , 
\end{array}
\end{equation}
where $\braket{\boldsymbol{r}_p^2}$ and $\braket{\boldsymbol{r}_n^2}$ is due to the finite size effect of proton and neutron, respectively, whereas $\braket{\boldsymbol{r}_{ls}^2}$ is the contribution of the relativistic spin-orbit component of the charge density~\eqref{zzz}. The Clebsch-Gordan coefficient $C^p_{T_\pi}$ is given in Eq.~\eqref{cg}. The proton pick-up spectroscopic factor $S_{k_\alpha,\pi}^i$ for a valence orbital $k_\alpha$ in the initial state $\ket{i}$ can be obtained by using the shell model. While the proton occupancy for a core orbital is taken as $(2j+1)$. More detail on the treatment of the core orbital contribution has been discussed in Ref.~\cite{XaNa2018}. 

Noticing that Eq.~\eqref{chrad} can be expressed in term of the proton density as usual, but the following modification must be introduced 
\begin{equation}
\rho_p(r) = \frac{1}{4\pi}\sum_{k_\alpha\pi} \left(C^p_{T_\pi}\right)^2 S_{k_\alpha,\pi}^i |R_{k_\alpha,p}^\pi(r)|^2\,. 
\end{equation}
If one substitutes this proton density into the mean field equations and solves them self-consistently, the result would be a combination of the shell model and HF method, and subsequently the radial wave functions can be improved. Similar technique has been employed in Ref.~\cite{BMH}. In the current work, we only use it for the charge radius calculation after the HF variation. 

The proton and neutron radii to be accounted with Eq.~\eqref{chrad} can be deduced from the effective electromagnetic form factor~\eqref{for}.  The final result is
\begin{equation}\label{nrad}
\displaystyle \braket{\boldsymbol{r}_q^2} = \frac{3}{2}\sum_i^{n_q} \left\{ a_q^i (r_q^i)^2+a_q^i\left[ \frac{1}{2}\left( \frac{\hbar}{mc}\right)^2-\frac{b^2}{A} \right]\right\}, 
\end{equation}
note that the Darwin-Foldy and COM corrections are also included in Eq.~\eqref{nrad}. The contribution from the spin-orbit component of the charge density has to be calculated numerically. Since this component is small, the classical method is applied for its evaluation: 
\begin{equation}
\displaystyle \braket{\boldsymbol{r}_{ls}^2} = {\int_0^\infty \rho_{ch}^{ls}(r)r^4dr}/{\int_0^\infty \rho_{ch}^{ls}(r)r^2dr}. 
\end{equation}

One can see that the inclusion of the CSB force and the use of the GGA in lieu of the Slater approximation does not affect the charge-radius calculation significantly. Moreover, although the contribution due to the suppression of isospin spuriosity can be as large as 0.12~fm (in the $A=10$ nuclei), the main features such as local variations are essentially unchanged. On the other hand, the deviations of the theoretical values from the experimental data are, in most cases, much larger than the experimental errors. In particular, with only few exceptions, the deviation for an odd-odd emitter with $N=Z$ is larger than that for its even-even partner. 

The correlation between $\delta_{RO}$ and charge radius which normally is proportional to the potential length, can be qualitatively understood from the study of the finite spherical square well. 
From the analytical solution for $l=0$ of this simplified model, a smaller radius yields lower eigenstates, making thus the radial wave functions to be more compressed in the direction towards the origin and less sensitive to the Coulomb potential which dominates the isospin-symmetry breaking at large separation distances. 

Subsequently, the corresponding value for $\delta_{RO}$ should also be smaller. Therefore, it is evident that, one of the reason for the discrepancy between the WS and Skyrme-HF calculations, especially the local effects, is the difference in the result they obtain for the charge radii. The correction values obtained with our HF protocol could be significantly improved if one is able to properly bring the calculated charge radii to match the measured values. 

\subsection{Standard model implication}\label{st}

We have further assessed all the five significant sets of calculated $\delta_{RO}$ values based on the CVC hypothesis. The corrected $\mathcal{F}t$ value was calculated for each transition using the expression~\eqref{Ft}. The data for $ft$ and the other theoretical corrections were taken from Ref.~\cite{HaTo2015}. To make our comparison easier, we assumed that each set of $\delta_{RO}$ values has the same uncertainties which were adopted from the calculation with WS potential in Ref.~\cite{NaXa2018}. These uncertainties were evaluated from the use of different shell-model effective interactions and from the experimental data on charge radii~\cite{AngMari2013,ToHa2002}. More details on the uncertainty determination for $\delta_{RO}$ were discussed in Ref.~\cite{XaNa2018}. 

The averaged $\mathcal{F}t$ value, and the normalized $\chi^2/\nu$ value which is the measure of scatters presented in the individual $\mathcal{F}t$ values, are listed for each calculation in Table~\ref{tab11}. Our method for data analysis is similar to that of the Particle Data Group~\cite{PDG}. The systematic uncertainties on $\delta_R'$ and $\delta_{NS}$ were evaluated using the method such as described in Ref.~\cite{HaTo2015}. If we assume that the CVC hypothesis is validated, the calculation using WS radial wave functions would be the best because it yields the smallest value for $\chi^2/\nu$. While the sets obtained with HF radial wave functions produce a remarkably larger value of both averaged $\mathcal{F}t$ and $\chi^2/\nu$. 
We notice that the the averaged $\mathcal{F}t$ values listed in Table~\ref{tab1} are much differed from the latest evaluation of Hardy and Towner~\cite{HaTo2015}, including that calculated with WS radial wave functions. 
The principal reason is that we have selected a smaller configuration space for some $sd$ and $fp$-shell nuclei as already mentioned in section~\ref{sm}.

\begin{table}[ht!]
\setlength{\extrarowheight}{0.08cm}
\caption{\label{tab11} The averaged $\mathcal{F}t$ value obtained from each set of calculated $\delta_{RO}$ values. $\chi^2/\nu$ is used to measure the scatters of the individual $\mathcal{F}t$ values from the averaged one. The scaling factor used in the evaluation of uncertainty is also listed.}
\begin{ruledtabular}
\begin{tabular}{cccc}
Calculations of $\delta_{RO}$	&	Averaged $\mathcal{F}t$~[sec.]	&	$\chi^2/\nu$	&	Scale	\\
\hline
SkM*	&	$3077.59(92)$	&	2.870	&	1.629	\\
SkM*-T	&	$3076.58(96)$	&	3.096	&	1.694	\\
VP-T	&	$3076.73(96)$	&	3.121	&	1.701	\\
FNS-T	&	$3076.56(96)$	&	3.113	&	1.699	\\
CSO-T	&	$3076.56(94)$	&	2.969	&	1.654	\\
GGA-T	&	$3076.22(97)$	&	3.160	&	1.713	\\
CIB-T	&	$3076.55(96)$	&	3.106	&	1.697	\\
CSB-T	&	$3075.10(92)$	&	2.869	&	1.625	\\
All-T	&	$3074.81(93)$	&	2.959	&	1.649	\\
WS	&	$3073.19(71)$	&	1.652	&	1.252 \\
\end{tabular}
\end{ruledtabular}
\end{table}

Another test of the correction values was performed using the experimentally determined $ft$ ratios of the mirror-pair transitions~\cite{Park2014}. 
Within the CVC validated ($\mathcal{F}t$ is nucleus-independent) the mentioned ratio is related to the theoretical corrections through the following expression, 
\begin{equation}\label{ratio}
\frac{ft^a}{ft^b} = 1 + ( \delta_R^{'b}-\delta_R^{'a} ) + ( \delta_{NS}^{b}-\delta_{NS}^{a} ) - ( \delta_C^{b}-\delta_C^{a} )
\end{equation}
where superscript $"a"$ denotes the decay of the even-even parent and $"b"$ denotes the mirror decay of the odd-odd parent. 
The expression in Eq.~\eqref{ratio} is advantageous because the theoretical uncertainties on the differences, $(\delta_R^{'b}-\delta_R^{'a})$, $(\delta_{NS}^{b}-\delta_{NS}^{a})$ and $(\delta_C^{b}-\delta_C^{a})$, are all less than on the individual correction terms themselves. 
More importantly, the ratio, ${ft^a}/{ft^b}$ is specifically sensitive to the local variations presented on $\delta_{RO}$. 

\begin{table*}[ht!]
\setlength{\extrarowheight}{0.08cm}
\caption{\label{tab44} Calculated ${ft^a}/{ft^b}$ values for the mirror pairs with $A = 26, 34, 38, 42, 50$, and $54$. The experimental data (EXP)~\cite{HaTo2015} for $A=26$, $34$ and $38$ are also listed in the last column.}
\begin{ruledtabular}
\begin{tabular}{cccccccccccc}
\multirow{2}{*}{$A$}        & \multicolumn{11}{c}{${ft^a}/{ft^b}$} \\
\cline{2-12}
	&	SkM*	&	SkM*-T	&	VP-T	&	FNS-T	&	CSO-T	&	GGA-T	&	CIB-T	&	CSB-T	&	All-T	&	WS	& EXP \\
\hline
26	&	1.00355	&	1.00304	&	1.00304	&	1.00304	&	1.00305	&	1.00303	&	1.00303	&	1.00312	&	1.00294	&	1.00488(46)	&	1.0048(19)	\\
34	&	1.00016	&	1.00005	&	1.00006	&	1.00006	&	1.00006	&	1.00005	&	1.00005	&	1.00001	&	1.00003	&	1.00177(64)	&	1.0028(10)	\\
38	&	0.99998	&	1.00021	&	1.0002	&	1.00022	&	1.00019	&	1.00022	&	1.00022	&	1.0002	&	1.00022	&	1.00167(57)	&	1.0037(20)	\\
42	&	1.00328	&	1.00288	&	1.00288	&	1.00288	&	1.00288	&	1.00287	&	1.00287	&	1.00281	&	1.0028	&	1.00523(88)	&		\\
46	&	1.00158	&	1.00158	&	1.00158	&	1.00158	&	1.00158	&	1.00158	&	1.00158	&	1.00158	&	1.00158	&	1.00158(167)	&		\\
50	&	1.00067	&	1.00033	&	1.00033	&	1.00034	&	1.00034	&	1.00034	&	1.00033	&	1.00025	&	1.00026	&	1.00184(85)	&		\\
54	&	1.00136	&	1.00101	&	1.00101	&	1.00101	&	1.00102	&	1.00101	&	1.00101	&	1.00096	&	1.00096	&	1.00266(149)	&		\\
\end{tabular}
\end{ruledtabular}
\end{table*}

The calculated ratios obtained for each calculation are shown in Table~\ref{tab44}. As can be seen, the WS set of correction values agrees very well with the available experimental data for the three mirror pairs with $A=26,34$ and $38$, while all calculations using Skyrme-HF radial wave functions differ from the data by almost two standard deviations. This is due to the positive sign that the Skyrme-HF calculations obtain for $(\delta_C^{b}-\delta_C^{a})$, which is unlike the result of WS. This strongly suggests that there is still some problems with the HF protocol adopted in the present work. 

\subsection{Impact of nucleon correlations}

In principle, a spherical HF method can only describe ground-state properties of spherical closed-shell nuclei. 
Its application to open-shell nuclei, including those considered in the present work, is more approximate due 
to the presence of nucleon correlations, as well as spontaneous symmetry breaking which is obviously a post-HF phenomenon. Since our many-body states are constructed from the shell-model diagonalization of an effective nucleon-nucleon Hamiltonian, containing all types of nucleonic correlations, we need only to worry about the appropriateness of the radial dependence of the spherically-symmetric single-particle wave functions. 
Therefore, it could be thought that, the remaining discrepancy between the WS and Skyrme-HF results may be rooted to the possible effects beyond the conventional HF minimization needed to match the experimental data on nucleon separation energies, used to refine the self-consistent potential. In order to move further with this argument, we decompose the binding energy of an open-shell nucleus as 
\begin{equation}\label{para}
E = E_{HF} + E_{PHF},
\end{equation}
where the first term, $E_{HF}$ can be found by using the spherical Skyrme-HF approximation and $E_{PHF}$ is what we call here a {\em post-HF} contribution. Working along $N=Z$ line, we notice that one particular effect which is definitely not taken into account by the conventional HF calculation is the Wigner contribution to the binding energies~\cite{wigner1,wigner2}. Therefore, we assume here that $E_{PHF}$ is dominated by the Wigner effect which is, in literature, parametrized as follows~\cite{MoNi1981,MoNi1995,Sat1997,Jensen1984,Zeldes1998,Kirson2008,YY}
\begin{equation}\label{ww}
E_{PHF} = W|N-Z|+d \, \delta_{NZ}^{\rm odd-odd}, 
\end{equation}
where $W$ and $d$ are smooth functions of mass and related each-other via $W\approx 3d/2$, and $\delta_{NZ}^{\rm odd-odd}=1$ for odd-odd $N=Z$ nuclei and zero for the others. 
Using Eq.~\eqref{para} and \eqref{ww} the Coulomb displacement energy between the initial and final states for a given mass can be expressed as, 
\begin{equation}\label{wig}
\displaystyle CDE = CDE_{HF} \pm 2d, 
\end{equation}
where the $+(-)$ sign corresponds to odd-odd(even-even) emitters. The $CDE$ is the total Coulomb displacement energy which can be matched with the experimental data, while $CDE_{HF}$ is the component which can be described using the HF approximation, and $\pm 2d$ is the Wigner or post-HF contribution. 

Eq.~\eqref{wig} demonstrates that the difference between the theoretical and the measured values of $CDE$ as shown in Fig.~\ref{figx}, is mainly due to the Wigner correlation. I.e., for an even-even emitter, the measured value is smaller than the HF value because the Wigner contribution to $CDE$ is negative. In contrast, the measured value is larger for an odd-odd emitter because the sign of the Wigner term is reversed to be positive. 

Next, we follow the usual adjustment procedure but use $CDE_{HF}$ extracted from Eq.~\eqref{wig} instead of the experimental data for $CDE$.  
Noticing, however, that within the standard value of the parameter $d$~\cite{YY}, we obtained unrealistic values for $\delta_{RO}$, namely it became too large for an even-even emitter and too small for an odd-odd emitter. In our opinion, this indicates that there should be other post-HF terms to weaken the Wigner contribution in Eq.~\eqref{wig}. To avoid this problem, we just treat $d$ as an adjustable parameter whose value yields the ratio ${ft^a}/{ft^b}$ to be the same as that obtained from the WS result which we expect to be closest to the real value. 

\begin{figure*}[htb]
\begin{center}
\includegraphics[]{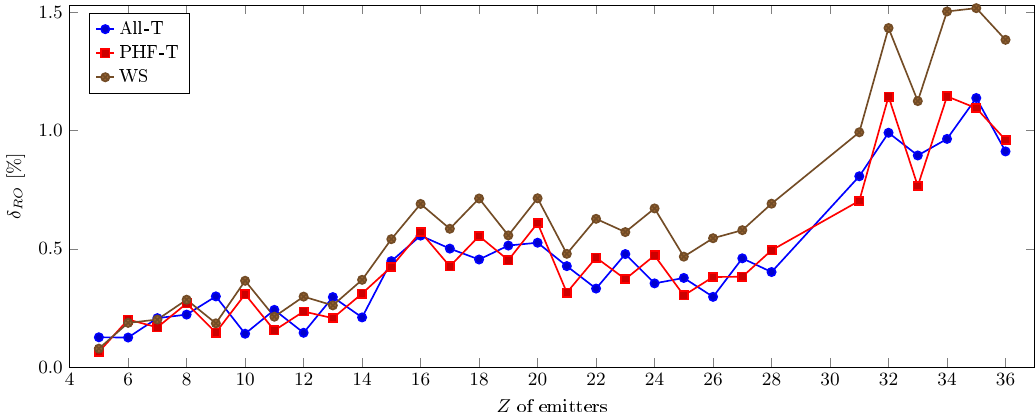}
\caption{\label{fig10}(Color online) Result for $\delta_{RO}$ after correction for the post-HF (PHF) contributions.}
\end{center}
\end{figure*}

As can be seen from Fig.~\ref{fig10}, the local variations on $\delta_{RO}$ are very sensitive to the post-HF contributions, with PHF subtracted from the experimental data for the Coulomb displacement energy, the Skyrme-HF result is much more similar to the WS result, except that the latter is still considerably larger in magnitude for medium to heavy mass regions. As a possible reason for this, there could be other post-HF contributions which are not properly absorbed in the last term of Eq.~\eqref{wig}. 

Although the relation between ${ft^a}/{ft^b}$ and the theoretical corrections, which we have used to fit the parameter $d$ was established based on the CVC hypothesis, it only constraints the magnitude of the difference $\delta_{RO}^b-\delta_{RO}^a$ for a given mirror pair of transitions. It is thus interesting to check for the ability of the new correction set in bringing the experimental $ft$ data into a single nucleus-independent $\mathcal{F}t$ value. Following the same procedure employed in subsection~\ref{st}, we obtained the averaged $\mathcal{F}t$ value of $3076.871\pm0.910$~sec. and the $\chi^2/\nu$ value of 2.628 which is still considerably larger than the WS result. 

\section{Summary and Conclusion}\label{con}

We have presented new calculations of the radial overlap correction required in the analysis of superallowed $0^+\to 0^+$ nuclear $\beta$ decays. The radial wave functions were obtained by solving the local equivalent potential constructed from a HF solution using the zero-range effective Skyrme interaction. Similar to the calculations of Ormand and Brown~\cite{OrBr1985,OrBr1989x,OrBr1995,OrBr1989}, the potential depth is scaled so that the energy eigenvalues match the experimental separation energies and ensures the correctness of the asymptotic radial wave functions. 
Several studies at the mean-field level have been carried out, including the assessment of the validity of the Slater approximation for the Coulomb exchange term and the effect of induced mass uncertainty due to the use of different center-of-mass corrections. We found that both of them are only a minor deficiency and have no significant impact on $\delta_{RO}$. 
Moreover, we proposed a simplified method for suppressing the spurious isospin mixing from the self-consistent HF solution. When it is applied, the $\delta_{RO}$ values are generally remarkably increased. However, there are few cases for which the suppression does not produce any visible effect or even decreases $\delta_{RO}$ due to the competition between the Coulomb and density-driven nuclear isovector terms of the mean-field potential. We suggest that it would be more accurate to build the local equivalent potential from an isospin-invariant and spuriosity-free HF solution using the constrained HF procedure as proposed in Ref.~\cite{BS}. 
We have also included the charge-independence-breaking and charge-symmetry-breaking forces complemented with the conventional Skyrme interaction. 
We found that the former force does not have any significant impact on the $\delta_{RO}$ values. However, the latter force increases considerably the mismatch between proton and neutron radial wave functions and subsequently increases the $\delta_{RO}$ values between 10 and 30~\%. The electromagnetic effects such as vacuum polarization and Coulomb spin-orbit all have a negligible contribution to $\delta_{RO}$. Similarly, the introduction of the charge form factors of nucleons does not enhance or reduce $\delta_{RO}$ significantly, however it has a non-negligible contribution to the charge radius of the emitters, especially the form factor of proton. While the replacement of the Slater approximation with the generalized gradient approximation leads to a further increase of $\delta_{RO}$ between 2 and 14~\%. 
Finally, our analysis of post-HF effects indicates that correlations beyond the HF approximation are important and should be taken into account. This may signify that direct application of spherical HF wave functions is not sufficient and further efforts on how to incorporate those correlations into the proposed framework should be pursued.
\begin{acknowledgments}
We are grateful to B.~A.~Brown for providing us with the NuShellX@MSU shell-model code. 
We are indebted to M. Bender for numerous discussions, enlightening explanations and, especially, for careful reading of the manuscript and many important remarks.
We thank also K.~Bennaceur for useful discussions and for his kind permission to use his code for solving spherical Hartree-Fock equations with 
the Skyrme and Gogny effective interactions. 
L.~Xayavong thanks CENBG for the hospitality and financial support during his 3-week stay. 
The work is supported by IN2P3/CNRS, France, in the framework of the ``Isospin-symmetry breaking'' and 
``Exotic nuclei, fundamental interactions and astrophysics'' Master projects. 
Large-scale calculations have been carried out at M\'esocentre de Calcul Intensif Aquitain (MCIA), Universit\'e de Bordeaux, France. 
\end{acknowledgments}

\bibliography{prc}


\appendix 

\section{Standard Skyrme-Hartree-Fock equations}\label{appA}
\subsection{Effective interactions}\label{a1}

The main part of the conventional effective Skyrme interaction~\cite{Bender,karim} is given by
\begin{equation}\label{sk}
\begin{array}{ll}
\displaystyle v_{Sk} &= \displaystyle t_0(1+x_0P_\sigma)\delta+\frac{1}{2}t_1(1+x_1P_\sigma)(\boldsymbol{k'}^2\delta+\delta\boldsymbol{k}^2) \\[0.18in]
&+ \displaystyle t_2(1+x_2P_\sigma) \boldsymbol{k'}\cdot\delta\boldsymbol{k}+\frac{1}{6}t_3(1+x_3P_\sigma)\rho^\gamma(\boldsymbol{R})\delta, 
\end{array}
\end{equation}
where $\delta=\delta(\boldsymbol{r}_i-\boldsymbol{r}_j)$, $\boldsymbol{k}=(1/2i)(\boldsymbol{\nabla}_i-\boldsymbol{\nabla}_j)$ is the relative momentum operator acting on the wave function to the right, and $\boldsymbol{k'}$ is the adjoint of $\boldsymbol{k}$. The $P_\sigma$ is the spin-exchange operator and $\boldsymbol{R}=(\boldsymbol{r}_i+\boldsymbol{r}_j)/2$. 

The spin-orbit part used in the present work is the one of Bell and Skyrme~\cite{BeSk1956}
\begin{equation}\label{so}
\displaystyle v_{so} = iW_0\boldsymbol{k'}\cdot\delta(\boldsymbol{\sigma}\times\boldsymbol{k}), 
\end{equation}
the $\boldsymbol{\sigma}=\boldsymbol{\sigma}_i+\boldsymbol{\sigma}_j$ is the sum of the Pauli spin matrices acting on spin variables
of particles $i$ and $j$, respectively. 
The parameters $x_0$, $x_1$, $x_2$, $x_3$, $t_0$, $t_1$, $t_2$, $t_3$, $W_0$ and $\gamma$ are adjustable constants. 

In addition, the two-body Coulomb interaction between charged particles can be written as 
\begin{equation}\label{coul}
\displaystyle v_{Coul} = \left(\frac12-t_{iz}\right)\left(\frac12-t_{jz}\right)\frac{e^2}{|\boldsymbol{r}_i-\boldsymbol{r}_j|}, 
\end{equation}
where $e$ is the elementary charge and $t_{iz}$($t_{jz}$) is the isospin projection of the particle $i$($j$). 

The extended interaction terms such as the charge-symmetry-breaking and charge-independence-breaking forces are discussed separately in section~\ref{isb}.

\subsection{Energy density functionals}\label{a2}

The energy density functional derived from the effective interactions listed in the previous subsection can be decomposed into three terms, 
\begin{equation}
\mathcal{H} = \mathcal{H}_{kin} + \mathcal{H}_{Sk} + \mathcal{H}_{Coul}, 
\end{equation}
where the first term on the right-hand-side arises from the kinetic part of the Hamiltonian, whereas the second and the third terms correspond to the main Skyrme (including spin-orbit) and the two-body Coulomb interactions, respectively. Some details on the derivation of $\mathcal{H}$ can be found in Ref.~\cite{Vautherin}. 

The kinetic energy density functional, $\mathcal{H}_{kin}$ is proportional to the total kinetic density $\tau$, namely 
\begin{equation}\label{ko}
\mathcal{H}_{kin} = \frac{\hbar^2}{2m}\tau, 
\end{equation}
where $\tau=\tau_p+\tau_n$ with $\tau_p$ and $\tau_n$ being defined in subsection~\ref{hfpo}. The mass $m$ is taken as $m=(m_p+m_n)/2$. 

Note that, in practice, Eq.~\eqref{ko} must be slightly modified to account for the center-of-mass correction~\cite{karim,Bender}. More details on this point are discussed in subsection~\ref{ccom}.

The main Skyrme functional, $\mathcal{H}_{Sk}$ derived from the sum of interactions, $v_{Sk}+v_{so}$ takes the following form, 
\begin{widetext}
\begin{equation}\label{hsk}
\begin{array}{ll}
\mathcal{H}_{Sk} &=\displaystyle\frac{t_0}{2}\left[ \left( 1 + \frac{x_0}{2} \right)\rho^2 - \left( x_0 + \frac{1}{2} \right)\sum_q \rho_q^2 \right] + \frac{t_1}{4} \left\{ \left( 1 + \frac{x_1}{2} \right)\left[ \rho\tau + \frac{3}{4}(\bm{\nabla}\rho)^2 \right] -\left( x_1+\frac{1}{2} \right)\sum_q\left[ \rho_q\tau_q + \frac{3}{4}(\bm{\nabla}\rho_q)^2 \right] \right\} \\[0.18in]
& +\displaystyle\frac{t_2}{4} \left\{ \left( 1 + \frac{x_2}{2} \right)\left[ \rho\tau - \frac{1}{4}(\bm{\nabla}\rho)^2 \right] + \left( x_2+\frac{1}{2} \right)\sum_q\left[ \rho_q\tau_q - \frac{1}{4}(\bm{\nabla}\rho_q)^2 \right] \right\} - \frac{1}{16}( t_1x_1 + t_2x_2 )J^2 + \frac{1}{16}( t_1-t_2 )\sum_q \bm{J_q}^2 \\[0.18in]
& +\displaystyle\frac{t_3}{12}\rho^\gamma \left[ \left( 1 + \frac{x_3}{2} \right)\rho^2 - \left( x_3 + \frac{1}{2} \right)\sum_q\rho_q^2 \right] + \frac{W_0}{2} \Big( \bm{J}\cdot\bm{\nabla}\rho + \sum_q \bm{J_q}\cdot\bm{\nabla}\rho_q \Big), 
\end{array}
\end{equation}
\end{widetext}
where $\bm{\nabla}$ denotes the gradient operator, $\rho_q$ is the particle density and $\bm{J_q}$ is the spin current vector~\cite{karim}, whereas $\rho=\rho_p+\rho_n$ and $\bm{J}=\bm{J}_p+\bm{J}_n$.

The direct part of the Coulomb functional, $\mathcal{H}_{Coul}^{dir}$ derived from the two-body Coulomb interaction~\eqref{coul} involves an integral over the coordinate space as given below, 
\begin{equation}
\displaystyle\mathcal{H}_{Coul}^{dir}(\bm{r}) = \frac{e^2}{2}\rho_{ch}(\bm{r})\int \frac{ \rho_{ch}(\bm{r}') }{ |\bm{r}-\bm{r}'| } d^3\bm{r}',  
\end{equation}
where $\rho_{ch}$ denotes the nuclear charge density which is usually approximated 
by the point-like proton density $\rho_p$. More details on $\rho_{ch}$ are described in subsection~\ref{cdens}. 

Unfortunately, its exchange counterpart, $\mathcal{H}_{Coul}^{exc}$ cannot be treated exactly without scarifying the simplicities that come with use of zero-range Skyrme interactions~\cite{Skalski,LeBloas}. In nearly all self-consistent Skyrme HF calculations, the Coulomb exchange functional is
accounted for within the local Slater approximation~\cite{Slater} which leads to the attractive contribution expressed below, 
\begin{equation}\label{hex}
\displaystyle\mathcal{H}_{Coul}^{ex} = - \frac{3}{4}e^2\left( \frac{3}{\pi} \right)^{1/3}\rho_{ch}^{4/3}. 
\end{equation}

The validity of the Slater approximation within the HF framework is discussed in subsection~\ref{slat}. More reliable method for treating the Coulomb exchange term, namely the generalized gradient approximation, is discussed in subsection~\ref{gen}. 

\subsection{Mean field potentials}\label{a3}

The effective mass, $m_q^*(r)$ appearing in Eq.~\eqref{u} and Eq.~\eqref{local} can be defined as the first derivative of the total energy density functional, $\mathcal{H}$ with respect to the kinetic density $\tau_q$. The resulting expression is following,
\begin{equation}\label{m*}
\begin{array}{ll}
\displaystyle \frac{\hbar^2}{2m_q^*} & = \displaystyle\frac{\partial \mathcal{H}}{\partial \tau_q} = \frac{\hbar^2}{2m} + \frac{1}{8}\left[ t_1\Big(2+x_1\Big)+t_2\Big(2+x_2\Big) \right]\rho  \\[0.18in]
&+ \displaystyle\frac{1}{8}\left[t_1\Big(1+2x_1\Big)+t_2\Big(1+2x_2\Big)\right]\rho_q. 
\end{array} 
\end{equation}

Likewise, the nuclear central potential can be defined as the first derivative of $\mathcal{H}$ with respect to the nucleon density $\rho_q$. The result is given below,
\begin{widetext}
\begin{equation}\label{e2.33}
\begin{array}{ll}
U_q & \displaystyle = \frac{\partial \mathcal{H}}{\partial \rho_q} = t_0\left[ \Big(1+\frac{x_0}{2}\Big)\rho - \Big(x_0+\frac{1}{2}\Big)\rho_q \right] + \frac{t_1}{4}\left[ \Big(1+\frac{x_1}{2}\Big)\Big(\tau-\frac{3}{2}\Delta\rho\Big)-\Big(x_1+\frac{1}{2}\Big)\Big(\tau_q-\frac{3}{2}\Delta\rho_q\Big) \right] \\[0.18in] 
& +\displaystyle\frac{t_2}{4}\left[ \Big(1+\frac{x_2}{2}\Big)\Big(\tau+\frac{1}{2}\Delta\rho\Big)+\Big(x_2+\frac{1}{2}\Big)\Big(\tau_q+\frac{1}{2}\Delta\rho_q\Big) \right] + \frac{t_3}{12}\Big[ \Big(1+\frac{x_3}{2}\Big)\Big(2+\gamma\Big)\rho^{\gamma+1} \\[0.18in]
&- \displaystyle\Big(x_3-\frac{1}{2}\Big)\Big(2\rho^\gamma\rho_q+\gamma\rho^{\gamma-1}\sum_{q'}\rho_{q'}^2 \Big) \Big] - \frac{W_0}{2}\Big[ \frac{1}{r} (J+J_q) + \frac{1}{2}\frac{d}{dr}(J+J_q) \Big] , 
\end{array}
\end{equation}
\end{widetext}
where the symbol $\Delta$ denotes the Laplacian operator and $J_q$ denotes the non-vanishing component of the spin current vector $\bm{J}_q$ when the spherical symmetry is assumed. The expression for $J_q$ is given in subsection~\ref{hfpo}.

The spin-orbit potential can be defined as the first derivative of $\mathcal{H}$ with respect to the density $J_q$. The resulting expression reads~\cite{karim},  
\begin{equation}\label{wform}
\begin{array}{ll}
W_q&= \displaystyle \frac{\partial \mathcal{H}}{\partial J_q} =-\frac{1}{8}(t_1x_1+t_2x_2)J \\[0.18in] 
& \displaystyle+\frac{1}{8}(t_1-t_2)J_q + \frac{W_0}{2}\frac{d}{dr}(\rho+\rho_q).  
\end{array}
\end{equation} 

Note that the first and second terms on the right-hand-side of Eq.~\eqref{wform} result from the tensor coupling with spin and gradient. They correspond to the $\propto J^2$ and $\propto J_q^2$ terms in Eq.~\eqref{hsk}, respectively. These terms were often omitted in the earlier studies, for example they were not included in the fit of the SkM* parametrization~\cite{skm*}. 

Similarly, the Coulomb potential can be defined as the first derivative of $\mathcal{H}$ with respect to the charge density $\rho_{ch}$. The direct part is obtained as, 
\begin{equation}\label{eoe}
V_{Coul}^{dir}(\bm{r}) = \frac{e^2}{2}\int \frac{ \rho_{ch}(\bm{r}') }{ |\bm{r}-\bm{r}'| } d^3\bm{r}', 
\end{equation}
which, within the spherical symmetry, can be reduced to the following radial form, 
\begin{equation}
\begin{array}{ll}\label{Dir}
\displaystyle V_{Coul}^{dir}(r) =& \displaystyle 4\pi e^2\left[ \frac{1}{r}\int_0^r dr'\rho_{ch}(r')r'^2-\int_0^r dr'\rho_{ch}(r')r' \right. \\[0.18in]
&+ \displaystyle \left. \int_r^\infty dr'\rho_{ch}(r')r' \right]. 
\end{array}
\end{equation}

Whereas the exchange counterpart of the Coulomb potential, derived using the approximation~\eqref{hex}, is found to be, 
\begin{equation}\label{Sl}
\displaystyle V_{Coul}^{ex} = -e^2\left[\frac{3}{\pi}\rho_{ch}\right]^{1/3}. 
\end{equation}

\section{Numerical results}

\begin{table*}[ht!]
\setlength{\extrarowheight}{0.08cm}
\caption{\label{tab1} Calculated charge radii of the 30 emitters in unit of fm. The 'SkM*' denotes the values obtained with Ormand-Brown protocol using the SkM* parametrization for the effective Skyrme interaction. The various electromagnetic corrections, such as those due to the vacuum polarization (VP), the finite size (FS) of nucleons, the Coulomb spin-orbit (CSO), the generalized gradient approximation (GGA) for the treatment of the Coulomb exchange term are separately accounted, the resulting charge radii values are given in the fourth, fifth, sixth and seventh columns, respectively. The addition of the charge-independence-breaking (CIB) and charge-symmetry-breaking (CSB) terms within the conventional Skyrme parametrization produces the charge radii values as given in the eighth and ninth columns, respectively. The 'All' denotes the calculation which includes all electromagnetic corrections and all additional interaction terms. The additional label 'T' indicates the application of the spurious isospin supression as detailed in subsection~\ref{fit}.
For comparison, the experimental (EXP) values are listed in the last column.}
\begin{ruledtabular}
\begin{tabular}{ccccccccccc}
\multirow{2}{*}{Emitter}   & \multicolumn{10}{c}{$\sqrt{\braket{\boldsymbol{r}_{ch}^2}}$~[fm]} \\
\cline{2-11}
	&	SkM*	&	SkM*-T	&	VP-T	&	FNS-T	&	CSO-T	&	GGA-T	&	CIB-T	&	CSB-T	&	All-T	&	EXP	\\
\hline
$^{10m}$B	&	2.323	&	2.442	&	2.436	&	2.517	&	2.436	&	2.438	&	2.437	&	2.439	&	2.522	&		\\
$^{10}$C	&	2.431	&	2.549	&	2.548	&	2.626	&	2.549	&	2.551	&	2.549	&	2.553	&	2.632	&	2.470(60)~\cite{ToHa2002}	\\
$^{14m}$N	&	2.533	&	2.602	&	2.601	&	2.689	&	2.601	&	2.602	&	2.602	&	2.605	&	2.694	&		\\
$^{14}$O	&	2.627	&	2.706	&	2.704	&	2.790	&	2.706	&	2.707	&	2.707	&	2.710	&	2.795	&	2.740(40)~\cite{ToHa2002}	\\
$^{18m}$F	&	2.778	&	2.808	&	2.807	&	2.895	&	2.808	&	2.808	&	2.809	&	2.811	&	2.900	&		\\
$^{18}$Ne	&	2.841	&	2.855	&	2.854	&	2.942	&	2.855	&	2.855	&	2.855	&	2.857	&	2.946	&	2.971(8)~\cite{AngMari2013}	\\
$^{22m}$Na	&	2.882	&	2.905	&	2.905	&	2.995	&	2.906	&	2.905	&	2.906	&	2.909	&	2.999	&		\\
$^{22}$Mg	&	2.933	&	2.944	&	2.944	&	3.033	&	2.945	&	2.945	&	2.945	&	2.948	&	3.037	&	3.069(5)~\cite{Bao}	\\
$^{26m}$Al	&	2.987	&	3.005	&	3.005	&	3.095	&	3.006	&	3.005	&	3.006	&	3.009	&	3.100	&		\\
$^{26}$Si	&	3.019	&	3.025	&	3.025	&	3.115	&	3.026	&	3.026	&	3.026	&	3.029	&	3.120	&	3.100(30)~\cite{ToHa2002}	\\
$^{30m}$P	&	3.052	&	3.067	&	3.067	&	3.158	&	3.067	&	3.067	&	3.068	&	3.071	&	3.162	&		\\
$^{30}$S	&	3.110	&	3.122	&	3.121	&	3.211	&	3.122	&	3.122	&	3.124	&	3.127	&	3.216	&	3.247(5)~\cite{Bao}	\\
$^{34}$Cl	&	3.173	&	3.184	&	3.184	&	3.274	&	3.183	&	3.184	&	3.185	&	3.188	&	3.278	&	3.350(15)~\cite{Bao}	\\
$^{34}$Ar	&	3.216	&	3.226	&	3.226	&	3.315	&	3.226	&	3.227	&	3.227	&	3.232	&	3.319	&	3.365(4)~\cite{AngMari2013}	\\
$^{38m}$K	&	3.365	&	3.370	&	3.370	&	3.455	&	3.368	&	3.370	&	3.371	&	3.374	&	3.458	&		\\
$^{38}$Ca	&	3.398	&	3.403	&	3.402	&	3.488	&	3.401	&	3.403	&	3.404	&	3.408	&	3.491	&	3.460(6)~\cite{Bao}	\\
$^{42}$Sc	&	3.452	&	3.451	&	3.450	&	3.536	&	3.451	&	3.451	&	3.452	&	3.454	&	3.541	&	3.570(24)~\cite{AngMari2013}	\\
$^{42}$Ti	&	3.485	&	3.472	&	3.472	&	3.557	&	3.471	&	3.472	&	3.473	&	3.476	&	3.562	&	3.596(11)~\cite{Bao}	\\
$^{46}$V	&	3.525	&	3.519	&	3.519	&	3.604	&	3.520	&	3.519	&	3.520	&	3.523	&	3.609	&	3.630(10)~\cite{Bao}	\\
$^{46}$Cr	&	3.555	&	3.538	&	3.537	&	3.622	&	3.539	&	3.538	&	3.539	&	3.542	&	3.627	&	3.678(10)~\cite{Bao}	\\
$^{50}$Mn	&	3.599	&	3.591	&	3.590	&	3.674	&	3.593	&	3.591	&	3.592	&	3.594	&	3.680	&	3.712(20)~\cite{AngMari2013}	\\
$^{50}$Fe	&	3.619	&	3.601	&	3.600	&	3.684	&	3.602	&	3.601	&	3.601	&	3.604	&	3.690	&	3.733(17)~\cite{Bao}	\\
$^{54}$Co	&	3.664	&	3.648	&	3.648	&	3.732	&	3.650	&	3.648	&	3.649	&	3.652	&	3.737	&	3.712(8)~\cite{Bao}	\\
$^{54}$Ni	&	3.685	&	3.663	&	3.662	&	3.745	&	3.665	&	3.663	&	3.664	&	3.666	&	3.751	&	3.752(10)~\cite{Bao}	\\
$^{62}$Ga	&	3.809	&	3.786	&	3.786	&	3.868	&	3.786	&	3.786	&	3.787	&	3.790	&	3.872	&	3.908(9)~\cite{Bao}	\\
$^{62}$Ge	&	3.840	&	3.813	&	3.812	&	3.894	&	3.813	&	3.813	&	3.814	&	3.817	&	3.898	&		\\
$^{66}$As	&	3.835	&	3.810	&	3.810	&	3.892	&	3.808	&	3.810	&	3.812	&	3.815	&	3.894	&	4.020(100)~\cite{ToHa2002}	\\
$^{66}$Se	&	3.866	&	3.840	&	3.839	&	3.920	&	3.838	&	3.840	&	3.841	&	3.845	&	3.923	&		\\
$^{70}$Br	&	3.936	&	3.907	&	3.906	&	3.986	&	3.904	&	3.907	&	3.908	&	3.911	&	3.988	&	4.100(100)~\cite{ToHa2002}	\\
$^{70}$Kr	&	3.952	&	3.922	&	3.921	&	4.001	&	3.919	&	3.922	&	3.923	&	3.926	&	4.003	&		\\
\end{tabular}
\end{ruledtabular}
\end{table*}

\begin{table*}[ht!]
\setlength{\extrarowheight}{0.08cm}
\caption{\label{tab3} Numerical values of $\delta_{RO}$ in \% unit for the 14 mirror pairs of superallowed $0^+\to 0^+$ nuclear $\beta$ decays. The 'SkM*' denotes the values obtained with Ormand-Brown protocol using the SkM* parametrization for the effective Skyrme interaction. The various electromagnetic corrections, such as those due to the vacuum polarization (VP), the finite size (FS) of nucleons, the Coulomb spin-orbit (CSO), the generalized gradient approximation (GGA) for the treatment of the Coulomb exchange term are separately accounted, the resulting $\delta_{RO}$ values are given in the fourth, fifth, sixth and seventh columns, respectively. The addition of the charge-independence-breaking (CIB) and charge-symmetry-breaking (CSB) terms within the conventional Skyrme parametrization produces the $\delta_{RO}$ values as given in the eighth and ninth columns, respectively. The 'All' denotes the calculation which includes all electromagnetic corrections and all additional interaction terms. The final $\delta_{RO}$ values after correcting for the post-Hartree-Fock (PHF) contribution are listed in the eleventh column. The additional label 'T' indicates the application of the spurious isospin supression as detailed in subsection~\ref{fit}. For comparison, the $\delta_{RO}$ values calculated with Woods-Saxon (WS) radial wave functions are also listed in the last column.}
\begin{ruledtabular}
\begin{tabular}{cccccccccccc}
\multirow{2}{*}{Emitter}               & \multicolumn{11}{c}{$\delta_{RO}$~[\%]} \\
\cline{2-12}
	&	SkM*	&	SkM*-T	&	VP-T	&	FNS-T	&	CSO-T	&	GGA-T	&	CIB-T	&	CSB-T	&	All-T	&	PHF-T	&	WS	\\
\hline
$^{10m}$B	&	0.017	&	0.088	&	0.086	&	0.09	&	0.088	&	0.1	&	0.087	&	0.114	&	0.126	&	0.064	&	0.078	\\
$^{10}$C	&	0.153	&	0.088	&	0.086	&	0.091	&	0.09	&	0.1	&	0.089	&	0.11	&	0.125	&	0.201	&	0.187	\\
$^{14m}$N	&	0.1	&	0.165	&	0.163	&	0.167	&	0.164	&	0.179	&	0.165	&	0.2	&	0.207	&	0.168	&	0.201	\\
$^{14}$O	&	0.182	&	0.178	&	0.175	&	0.181	&	0.176	&	0.192	&	0.179	&	0.214	&	0.222	&	0.269	&	0.285	\\
$^{18m}$F	&	0.18	&	0.247	&	0.244	&	0.248	&	0.248	&	0.261	&	0.248	&	0.286	&	0.299	&	0.146	&	0.185	\\
$^{18}$Ne	&	0.146	&	0.111	&	0.108	&	0.112	&	0.112	&	0.121	&	0.111	&	0.13	&	0.141	&	0.308	&	0.365	\\
$^{22m}$Na	&	0.168	&	0.194	&	0.191	&	0.195	&	0.196	&	0.205	&	0.194	&	0.232	&	0.242	&	0.155	&	0.213	\\
$^{22}$Mg	&	0.137	&	0.107	&	0.104	&	0.108	&	0.109	&	0.114	&	0.106	&	0.136	&	0.145	&	0.235	&	0.298	\\
$^{26m}$Al	&	0.207	&	0.243	&	0.24	&	0.244	&	0.246	&	0.253	&	0.244	&	0.268	&	0.296	&	0.207	&	0.261	\\
$^{26}$Si	&	0.182	&	0.167	&	0.164	&	0.168	&	0.171	&	0.176	&	0.167	&	0.2	&	0.21	&	0.311	&	0.369	\\
$^{30m}$P	&	0.342	&	0.383	&	0.377	&	0.385	&	0.382	&	0.4	&	0.383	&	0.438	&	0.448	&	0.424	&	0.541	\\
$^{30}$S	&	0.391	&	0.479	&	0.473	&	0.478	&	0.48	&	0.496	&	0.485	&	0.547	&	0.556	&	0.572	&	0.690	\\
$^{34}$Cl	&	0.392	&	0.441	&	0.434	&	0.441	&	0.435	&	0.456	&	0.442	&	0.504	&	0.501	&	0.426	&	0.585	\\
$^{34}$Ar	&	0.359	&	0.397	&	0.391	&	0.398	&	0.392	&	0.412	&	0.398	&	0.456	&	0.455	&	0.554	&	0.713	\\
$^{38m}$K	&	0.41	&	0.457	&	0.451	&	0.457	&	0.449	&	0.472	&	0.458	&	0.522	&	0.514	&	0.452	&	0.557	\\
$^{38}$Ca	&	0.398	&	0.468	&	0.461	&	0.469	&	0.458	&	0.484	&	0.47	&	0.532	&	0.526	&	0.609	&	0.714	\\
$^{42}$Sc	&	0.339	&	0.366	&	0.361	&	0.367	&	0.371	&	0.377	&	0.367	&	0.417	&	0.427	&	0.315	&	0.479	\\
$^{42}$Ti	&	0.292	&	0.279	&	0.274	&	0.28	&	0.284	&	0.289	&	0.279	&	0.323	&	0.332	&	0.463	&	0.627	\\
$^{46}$V	&	0.379	&	0.413	&	0.407	&	0.413	&	0.418	&	0.424	&	0.414	&	0.469	&	0.478	&	0.372	&	0.571	\\
$^{46}$Cr	&	0.321	&	0.299	&	0.293	&	0.299	&	0.304	&	0.308	&	0.299	&	0.347	&	0.354	&	0.472	&	0.671	\\
$^{50}$Mn	&	0.297	&	0.317	&	0.312	&	0.316	&	0.325	&	0.323	&	0.317	&	0.367	&	0.377	&	0.303	&	0.467	\\
$^{50}$Fe	&	0.258	&	0.244	&	0.239	&	0.244	&	0.253	&	0.251	&	0.244	&	0.286	&	0.297	&	0.381	&	0.545	\\
$^{54}$Co	&	0.375	&	0.39	&	0.384	&	0.39	&	0.401	&	0.399	&	0.391	&	0.447	&	0.46	&	0.382	&	0.579	\\
$^{54}$Ni	&	0.357	&	0.337	&	0.331	&	0.337	&	0.349	&	0.346	&	0.338	&	0.389	&	0.402	&	0.494	&	0.691	\\
$^{62}$Ga	&	0.64	&	0.729	&	0.718	&	0.727	&	0.722	&	0.745	&	0.73	&	0.817	&	0.807	&	0.703	&	0.993	\\
$^{62}$Ge	&	0.824	&	0.907	&	0.894	&	0.905	&	0.901	&	0.927	&	0.927	&	1	&	0.991	&	1.144	&	1.434	\\
$^{66}$As	&	0.713	&	0.817	&	0.804	&	0.814	&	0.808	&	0.833	&	0.819	&	0.91	&	0.895	&	0.766	&	1.125	\\
$^{66}$Se	&	0.811	&	0.887	&	0.873	&	0.885	&	0.879	&	0.906	&	0.889	&	0.978	&	0.965	&	1.145	&	1.504	\\
$^{70}$Br	&	0.938	&	1.043	&	1.029	&	1.04	&	1.042	&	1.061	&	1.046	&	1.146	&	1.138	&	1.095	&	1.518	\\
$^{70}$Kr	&	0.685	&	0.826	&	0.812	&	0.823	&	0.825	&	0.842	&	0.827	&	0.918	&	0.912	&	0.961	&	1.384	\\
\end{tabular}
\end{ruledtabular}
\end{table*}

\end{document}